\documentclass[twocolumn,times]{aastex63} 

\usepackage{amssymb}
\usepackage{amsmath}
\usepackage{xspace}
\usepackage{epstopdf}
\usepackage[english]{babel}
\usepackage{mathrsfs}
\usepackage{mathtools}
\usepackage{graphicx}
\usepackage{listings}
\usepackage{xcolor}
\usepackage{array}
\usepackage{listings}

\newcolumntype{L}[1]{>{\raggedright\arraybackslash}p{#1}}
\newcolumntype{C}[1]{>{\centering\arraybackslash}p{#1}}
\newcolumntype{R}[1]{>{\raggedleft\arraybackslash}p{#1}}

\definecolor{dkgreen}{rgb}{0,0.6,0}
\definecolor{gray}{rgb}{0.5,0.5,0.5}
\definecolor{mauve}{rgb}{0.58,0,0.82}
\definecolor{red}{rgb}{1,0,0}

\newcommand{\nhosts}{$86$}
\newcommand{\nplanets}{$103$}

\newcommand{\equal}{$=$}

\newcommand{\gaia}{\textit{Gaia}}
\newcommand{\jwst}{\textit{JWST}}
\newcommand{\kepler}{\textit{Kepler}}

\newcommand{\feh}{$\rm{[Fe/H]}$}
\newcommand{\logg}{$\log g$}
\newcommand{\logrhk}{$\log R_{HK}^\prime$}

\newcommand{\mstar}{$M_{\star}$}
\newcommand{\msun}{$M_{\odot}$}

\newcommand{\rstar}{$R_{\star}$}

\newcommand{\teff}{$\rm T_{eff}$}
\newcommand{\vsini}{$v\sin i$}

\newcommand{\ator}{$a/R_{\star}$}

\newcommand{\radiuse}{$R_{\oplus}$}

\newcommand{\radiusp}{$R_{\mathrm{p}}$}

\newcommand{\sinc}{$\rm S_{inc}$}
\newcommand{\since}{$\rm S_{\oplus}$}

\newcommand{\nobs}{$\mathrm{N_{obs}}$}
\newcommand{\sigact}{$\sigma_{\mathrm{act}}$}

\newcommand{\sigrot}{$\sigma_{\mathrm{rot}}$}
\newcommand{\sigrv}{$\sigma_{\mathrm{RV}}$}
\newcommand{\sigtot}{$\sigma_{\mathrm{tot}}$}
\newcommand{\snr}{$\rm S/N$}
\newcommand{\tfive}{$t_\mathrm{5\sigma,HIRES}$}
\newcommand{\tten}{$t_\mathrm{10\sigma,HIRES}$}

\newcommand{\kms}{$\rm km\, s^{-1}$}
\newcommand{\ms}{$\rm m\, s^{-1}$}

\newcommand{\rom}[1]{{\scshape \romannumeral #1}}

\newcommand{\plotcolor}{orange}

\shorttitle{The TESS-Keck Survey}
\shortauthors{Chontos et al.}

\begin{document}

\title{The TESS-Keck Survey: Science Goals and Target Selection}

\author[0000-0003-1125-2564]{Ashley Chontos}
\altaffiliation{NSF Graduate Research Fellow}
\affiliation{Institute for Astronomy, University of Hawai`i, 2680 Woodlawn Drive, Honolulu, HI 96822, USA}


\author[0000-0001-8898-8284]{Joseph M. Akana Murphy}
\altaffiliation{NSF Graduate Research Fellow}
\affiliation{Department of Astronomy and Astrophysics, University of California, Santa Cruz, CA 95064, USA}

\author[0000-0003-2562-9043]{Mason G MacDougall}
\affiliation{Department of Physics \& Astronomy, University of California Los Angeles, Los Angeles, CA 90095, USA}

\author[0000-0002-3551-279X]{Tara Fetherolf}
\affiliation{Department of Earth and Planetary Sciences, University of California, Riverside, CA 92521, USA}

\author[0000-0002-4290-6826]{Judah Van Zandt}
\affiliation{Department of Physics \& Astronomy, University of California Los Angeles, Los Angeles, CA 90095, USA}

\author[0000-0003-3856-3143]{Ryan A. Rubenzahl}
\altaffiliation{NSF Graduate Research Fellow}
\affiliation{Department of Astronomy, California Institute of Technology, Pasadena, CA 91125, USA}

\author[0000-0001-7708-2364]{Corey Beard}
\affiliation{Department of Physics \& Astronomy, University of California Irvine, Irvine, CA 92697, USA}

\author[0000-0001-8832-4488]{Daniel Huber}
\affiliation{Institute for Astronomy, University of Hawai`i, 2680 Woodlawn Drive, Honolulu, HI 96822, USA}


\author[0000-0002-7030-9519]{Natalie M. Batalha}
\affiliation{Department of Astronomy and Astrophysics, University of California, Santa Cruz, CA 95060, USA}

\author{Ian J. M. Crossfield}
\affiliation{Department of Physics \& Astronomy, University of Kansas, 1082 Malott,1251 Wescoe Hall Dr., Lawrence, KS 66045, USA}

\author[0000-0001-8189-0233]{Courtney D. Dressing}
\affiliation{{Department of Astronomy,  University of California Berkeley, Berkeley CA 94720, USA}}

\author[0000-0003-3504-5316]{Benjamin Fulton}
\affiliation{NASA Exoplanet Science Institute/Caltech-IPAC, MC 314-6, 1200 E California Blvd, Pasadena, CA 91125, USA}

\author[0000-0001-8638-0320]{Andrew W. Howard}
\affiliation{Department of Astronomy, California Institute of Technology, Pasadena, CA 91125, USA}

\author[0000-0002-0531-1073]{Howard Isaacson}
\affiliation{{Department of Astronomy,  University of California Berkeley, Berkeley CA 94720, USA}}
\affiliation{Centre for Astrophysics, University of Southern Queensland, Toowoomba, QLD, Australia}

\author[0000-0002-7084-0529]{Stephen R. Kane}
\affiliation{Department of Earth and Planetary Sciences, University of California, Riverside, CA 92521, USA}

\author[0000-0003-0967-2893]{Erik A. Petigura}
\affiliation{Department of Physics \& Astronomy, University of California Los Angeles, Los Angeles, CA 90095, USA}

\author[0000-0003-0149-9678]{Paul Robertson}
\affiliation{Department of Physics \& Astronomy, University of California Irvine, Irvine, CA 92697, USA}

\author[0000-0001-8127-5775]{Arpita Roy}
\affiliation{Space Telescope Science Institute, 3700 San Martin Drive, Baltimore, MD 21218, USA}
\affiliation{Department of Physics and Astronomy, Johns Hopkins University, 3400 N Charles St, Baltimore, MD 21218, USA}

\author[0000-0002-3725-3058]{Lauren M. Weiss}
\affiliation{Institute for Astronomy, University of Hawai`i, 2680 Woodlawn Drive, Honolulu, HI 96822, USA}


\author[0000-0003-0012-9093]{Aida Behmard}
\altaffiliation{NSF Graduate Research Fellow}
\affiliation{Division of Geological and Planetary Science, California Institute of Technology, Pasadena, CA 91125, USA}

\author[0000-0002-8958-0683]{Fei Dai}
\affiliation{Division of Geological and Planetary Sciences
1200 E California Blvd, Pasadena, CA, 91125, USA}

\author[0000-0002-4297-5506]{Paul A. Dalba}
\altaffiliation{NSF Astronomy and Astrophysics Postdoctoral Fellow}
\affiliation{Department of Earth and Planetary Sciences, University of California, Riverside, CA 92521, USA}

\author[0000-0002-8965-3969]{Steven Giacalone}
\affiliation{Department of Astronomy, University of California Berkeley, Berkeley, CA 94720, USA}

\author[0000-0002-0139-4756]{Michelle L. Hill}
\affiliation{Department of Earth and Planetary Sciences, University of California, Riverside, CA 92521, USA}

\author[0000-0001-8342-7736]{Jack Lubin}
\affiliation{Department of Physics \& Astronomy, University of California Irvine, Irvine, CA 92697, USA}

\author{Andrew Mayo}
\altaffiliation{NSF Graduate Research Fellow}
\affiliation{Department of Astronomy, University of California Berkeley, Berkeley, CA 94720, USA}

\author[0000-0003-4603-556X]{Teo Mo\v{c}nik}
\affiliation{Gemini Observatory/NSF's NOIRLab, 670 N. A`ohoku Place, Hilo, HI 96720, USA}

\author{Alex S. Polanski}
\affiliation{Department of Physics \& Astronomy, University of Kansas, 1082 Malott,1251 Wescoe Hall Dr., Lawrence, KS 66045, USA}

\author[0000-0001-8391-5182]{Lee J.\ Rosenthal}
\affiliation{Department of Astronomy, California Institute of Technology, Pasadena, CA 91125, USA}

\author[0000-0003-3623-7280]{Nicholas Scarsdale}
\affiliation{Department of Astronomy and Astrophysics, University of California, Santa Cruz, CA 95060, USA}

\author{Emma V. Turtelboom}
\affiliation{{Department of Astronomy,  University of California Berkeley, Berkeley CA 94720, USA}}
\correspondingauthor{Ashley Chontos}
\email{achontos@hawaii.edu}

\begin{abstract}
\noindent Space-based transit missions such as \kepler\ and TESS have demonstrated that planets are ubiquitous. However, the success of these missions heavily depends on ground-based radial velocity (RV) surveys, which combined with transit photometry can yield bulk densities and orbital properties. While most \kepler\ host stars are too faint for detailed follow-up observations, TESS is detecting planets orbiting nearby bright stars that are more amenable to RV characterization. Here we introduce the TESS-Keck Survey (TKS), an RV program using $\sim$100 nights on Keck/HIRES to study exoplanets identified by TESS. The primary survey aims are investigating the link between stellar properties and the compositions of small planets; studying how the diversity of system architectures depends on dynamical configurations or planet multiplicity; identifying prime candidates for atmospheric studies with \jwst; and understanding the role of stellar evolution in shaping planetary systems. We present a fully-automated target selection algorithm, which yielded \nplanets\ planets in \nhosts\ systems for the final TKS sample. Most TKS hosts are inactive, solar-like, main-sequence stars ($4500$ K $\leq$ \teff\ $< 6000$ K) at a wide range of metallicities. The selected TKS sample contains 71 small planets (\radiusp\ $\leq$ 4 \radiuse), 11 systems with multiple transiting candidates, 6 sub-day period planets and 3 planets that are in or near the habitable zone (\sinc $\leq$ 10 \since) of their host star. The target selection described here will facilitate the comparison of measured planet masses, densities, and eccentricities to  predictions from planet population models. Our target selection software is publicly available\footnote{\url{https://github.com/ashleychontos/sort-a-survey}} and can be adapted for any survey which requires a balance of multiple science interests within a given telescope allocation.
\end{abstract}


\section{Introduction} \label{sec:intro}

Measurements of planet sizes in combination with masses, via the transit and radial velocity (RV) methods, continues to be the most fruitful synergy for investigating the properties (in particular the bulk compositions) of exoplanets. Early landmark discoveries using both transits and RVs included the first rocky exoplanets, Kepler 10-b \citep{batalha2011} and CoRoT-7-b \citep{leger2011}, as well as the density transition at 1.5 \radiuse\ \citep{weiss2014}, separating primarily rocky planets from the lower density, volatile-rich sub-Neptune-sized  planets \citep{rogers2015}. 

\kepler\ identified thousands of new transiting candidates \citep{borucki2011,borucki2011a,batalha2013,burke2014,rowe2015,mullally2015,coughlin2016,thompson2018}, but most targets were faint and therefore had limited constraints from spectroscopic RV surveys. Fortunately this has been mitigated with TESS \citep[Transiting Exoplanet Survey Satellite;][]{ricker2015}, a nearly all-sky survey looking for transiting planets orbiting bright nearby stars. The first TESS discovery of a rocky super-Earth orbiting the naked-eye star $\pi$ Men \citep{huang2018} was a prime example of the goals from the new mission.

A fundamental goal of exoplanet demographics is to compare observed populations of exoplanet properties to population synthesis models in order to inform planet formation theories \citep{mordasini2018}. This was accomplished with \kepler\ for planet radii \citep{mulders2019} and is now becoming possible for densities with TESS, but the selection function for RV follow-up is typically much more complex than for transit surveys. For example, RV surveys frequently drop stars with rapid rotation or increased stellar activity, which is typically done on a case-by-case basis and is thus difficult to reproduce in later studies.

Large exoplanet surveys such as TESS provide the statistical insights required to test different formation and evolution theories to observed planet distributions, but hinges on the ability to reproduce RV survey selection biases. Therefore, a critical ingredient for the realization of these surveys is understanding the process for which targets were initially selected. Indeed, recent ground-based TESS follow-up programs such as the Magellan-TESS Survey \citep{teske2020} have begun to describe target selection functions, providing a pathway to properly correct for survey biases or incompleteness.

Here we introduce the TESS-Keck Survey (TKS), a collaboration between the California Institute of Technology, the University of California (Berkeley, Irvine, Los Angeles, Riverside, Santa Cruz), the University of Hawai'i, the University of Kansas, NASA, the NASA Exoplanet Science Institute and the W. M. Keck Observatory. The survey is being conducted using the High-Resolution Spectrograph (HIRES), which is mounted on the Keck I telescope at the W. M. Keck Observatory on the summit of Maunakea in Hawai'i. Building on the legacies of \kepler\ and K2, TKS will leverage the new population of transiting planets orbiting nearby, bright stars to address major outstanding questions in exoplanet astronomy. We discuss the detailed vetting and target selection steps taken to converge on a definitive target list, including a fully-automated target selection algorithm. We conclude by presenting our complete target sample and summarize some of the general population characteristics.

\section{Survey Description}

\subsection{TKS Science} \label{sec:science}

TKS is structured around several science goals related to the compositions, architectures and atmospheres of exoplanets (see Table \ref{tab:themes}). The following subsections provide a brief review of the science motivations for the survey.


\begin{table*}
\begin{center}
\caption{TESS-Keck Survey Science Summary}
\begin{tabular}{L{1.8cm} L{2.8cm} L{1.0cm} L{10.5cm}}
\hline
\noalign{\smallskip}
\textbf{Theme} & \textbf{Science Case} & \textbf{ID$^{*}$} & \textbf{Description} \\
\noalign{\smallskip}
\hline
\noalign{\smallskip}
Bulk Compositions & Planet Radius Gap & 1A & Probing compositions across the planet radius gap to constrain the physical mechanism(s) causing the bimodal radius distribution  \\
\noalign{\smallskip}
& Stellar Flux \& Gaseous Envelopes & 1B & Analyzing the diversity of gaseous envelopes and their dependence on properties like stellar mass, insolation flux, activity level, and other properties \\
\noalign{\smallskip}
& Ultra-Short Period Planets & 1C & Using ultra-short period planet compositions as a window into the refractory cores of small planets \\
\noalign{\smallskip}
& Habitable-Zone Planets & 1D & Identifying and characterizing planets orbiting in the habitable zones of their host star \\
\noalign{\smallskip}
& Planet-Star Correlations & 1E & Exploring dependencies of bulk planet properties with different stellar properties like \mstar, \feh, age, etc. \\
\noalign{\smallskip}
\hline
\noalign{\smallskip}
Architectures \& Dynamics & Distant Giants & 2A & Understanding the occurrence and connection between close-in small planets and distant giant planets \\
\noalign{\smallskip}
& Eccentricities & 2Bi & Characterizing the eccentricities of sub-Jovian planets to elucidate the possible formation and/or evolution pathways for dynamically hot planets \\
\noalign{\smallskip}
& Obliquities & 2Bii & Measuring spin-orbit (mis)alignments of previously unexplored parameter spaces to trace formation histories and past dynamical interactions \\
\noalign{\smallskip}
& Multis & 2C &  Examining the diversity and/or uniformity of properties for planets in multi-planet systems \\
\noalign{\smallskip}
\hline
\noalign{\smallskip}
Atmospheres & & 3 & Identifying interesting planets amenable to atmospheric characterization using transmission and/or eclipse spectroscopy \\
\noalign{\smallskip}
\hline
\noalign{\smallskip}
Evolved & & 4 & Investigating the role and degree to which stellar evolution plays in shaping post-main sequence planetary systems \\
\noalign{\smallskip}
\hline
\noalign{\smallskip}
Technical Outcomes & Planet Host Properties & TA & Homogeneously determining fundamental stellar properties derived from spectroscopy, asteroseismology, and astrometry \\
\noalign{\smallskip}
& Astrophysical Doppler Noise & TB & Studying the relationship between Doppler jitter with various stellar astrophysical processes \\
\noalign{\smallskip}
\hline
\multicolumn{4}{l}{{\sc \textbf{Note ---}}} \\
\multicolumn{4}{l}{$^{*}$ The keys in this table are used in the final TKS sample in Tables \ref{tab:sample1}-\ref{tab:sample3} to identify which science case(s) each target will address.} \\
\end{tabular}
\label{tab:themes}
\end{center}
\end{table*}


\subsubsection{Bulk Compositions} \label{sec:SC1}

One of the most significant exoplanet discoveries by \kepler\ was the prevalence of planets between the sizes of Earth and Neptune \citep{howard2012,fressin2013,petigura2013}. More recently, refined radius measurements identified a valley in the planet distribution at a radius of $\sim$1.8 \radiuse\ \citep{fulton2017,fulton2018,vaneylen2018}. 

The dominant theory for explaining the radius valley is that the two populations originated from a single continuous distribution and an outcome sculpted by post-formation pathways. An example is photoevaporation, a process that strips away the atmosphere for less massive planets that receive high-energy incident flux from their host star \citep{owen2013,owen2017}. Another mechanism is core-powered mass-loss, which is also able to reproduce the observed planet radius distribution based solely on the internal cooling of a planet \citep{ginzburg2018,gupta2019,gupta2020}. However, a major caveat is that the simulated planet populations used to test these theories require assumptions about the underlying mass distribution, which is degenerate with core compositions. 

Multi-planet transiting systems with planets that straddle the radius valley are benchmarks for discerning between these theories. \citet{owen2020} used 104 planets in 73 systems and found that only two planets were significantly ($>3\sigma$) inconsistent with the photoevaporation model. However, all planets used in the analysis were faint \kepler\ systems, where less than half of the population had actual mass constraints. TESS planets are more amenable to RV characterization and are therefore an ideal population to test these theories. A recent example was demonstrated in \citet{cloutier2020b} for TOI-732, an M-dwarf with a pair of planets that straddle the gap, but more systems around a larger diversity of host star spectral types are needed.

\begin{figure*}
\includegraphics[width=\linewidth]{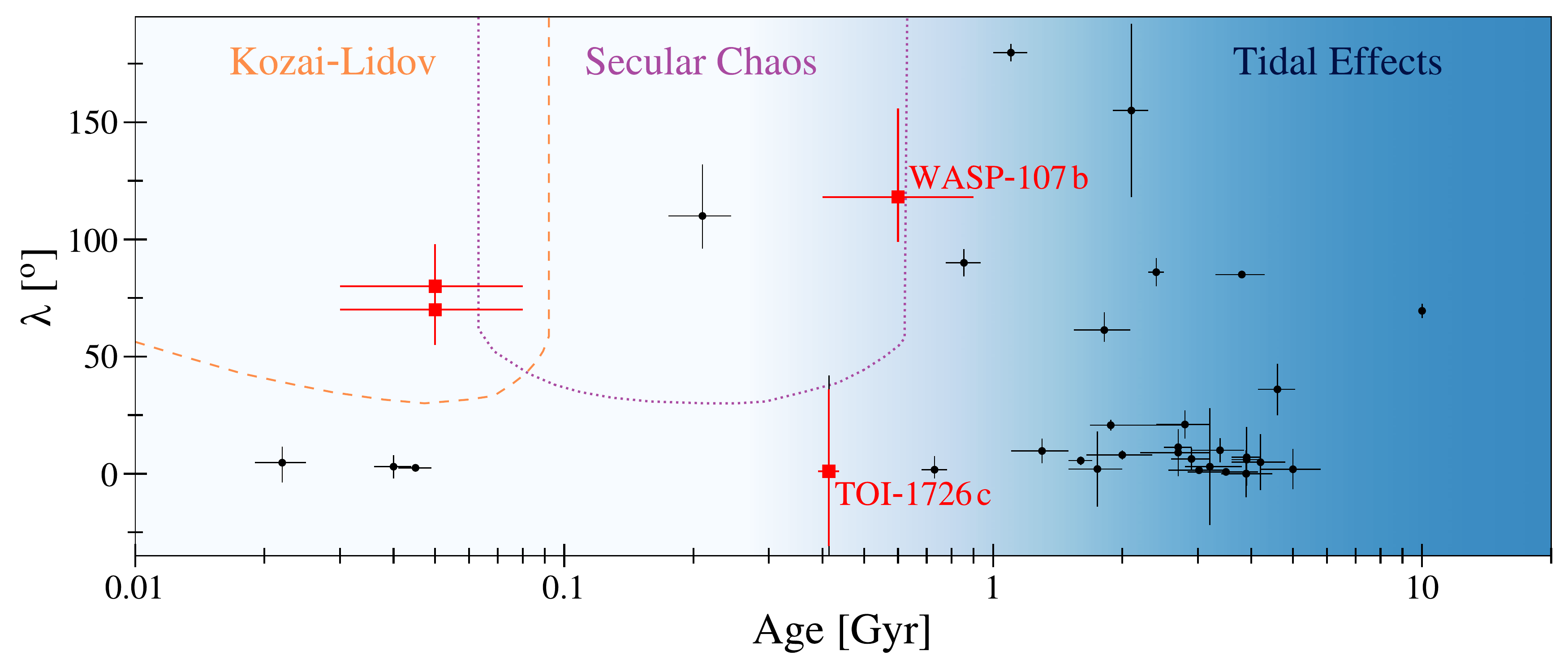}
\caption{Stellar obliquity versus host star age for confirmed planetary systems (Figure adapted from \citealt{zhou2020}). Tidal realignment and other obliquity-exciting mechanisms such as Kozai-Lidov or secular processes operate on different timescales. Young systems with well-known ages may elucidate the relative importance of these mechanisms. TOI-1726 c \citep{dai2020}, WASP-107 b \citep{rubenzahl2021} and simulated upcoming TKS measurements (red) add crucial points
to distinguish various obliquity-exciting scenarios.}
\label{fig:obliquity}
\end{figure*}

TKS will measure precise ($>$5$\sigma$) masses of small planets ($<4$ \radiuse) in various environments to address some of these questions. In particular, TKS aims to probe planet compositions across the radius gap in order to constrain the underlying physical mechanism(s), including timescales for which the bimodal distribution becomes more distinct. Additionally, by targeting a range of small planets ($1$ \radiuse\ $<$ \radiusp\ $< 4$ \radiuse) at various incident fluxes, TKS can specifically test the photoevaporation hypothesis. The compositions for ultra-short period planets (P$<$1 day) will be used as a window into the refractory core of small planets \citep{dai2021}. TKS will also measure masses for cooler planets in order to identify any that could have possible Earth-like compositions, especially for planets that are amenable to subsequent atmospheric characterization. Finally, TKS will make use of the entire population of small planets with precisely measured masses to investigate any dependencies or correlations of bulk planet properties with stellar properties.

\subsubsection{System Architectures \& Dynamics} \label{sec:SC2}

A striking feature of the solar system is its dynamically cool architecture; i.e., planets have nearly circular, nearly co-planar orbits that are well-aligned with the rotation axis of the Sun. 

On the contrary, large-scale exoplanet surveys have since introduced us to planets in a rich diversity of system architectures and dynamical configurations. For instance, it is now believed that orbital excitation is a common outcome of various planet formation channels \citep{ford2008} or dynamical perturbation scenarios \citep{goldreich2014}. While planet detections from radial velocity surveys have yielded well-constrained eccentricity measurements for many giant planets, sub-Jovian planet dynamics and dynamical histories remain elusive. Current observations point towards smaller planets having lower eccentricities \citep{vaneylen2015,vaneylen2019}, but the number of such systems with precise eccentricities is low. 

Another dynamical property is the obliquity, or the angle between the orbital plane of the system with respect to the stellar rotation axis. There are many proposed mechanisms for spin-orbit misalignments in planetary systems that operate on different timescales (Figure \ref{fig:obliquity}). Some processes tilt planet orbits during the disk-hosting stage \citep[$\sim$3 Myr,][]{batygin2011}; while the Kozai-Lidov mechanism operates typically on tens of Myr timescales, depending on the system configuration \citep{fabrycky2007}. Another prominent theory occurs through secular interactions between planets and is expected to happen in hundreds of Myr \citep{wu2011}. Therefore, obliquity measurements of 
stars with well-known ages could distinguish these theories. Young planets, especially those have established cluster membership, have the best age estimate and provide the strongest constraint on various obliquity-exciting theories.

TKS will shed light on these questions by measuring eccentricities and constraining obliquities around stars of various ages. Early TKS studies have already yielded obliquities for two new systems, including a small planet in a multi-planet transiting system in the Ursa Major Moving Group \citep{dai2020} and the nearly pole-on, superpuff WASP-107 b \citep{rubenzahl2021}. Additionally, TKS will increase the number of well-studied eccentric sub-Jovians by selecting high-probability eccentric candidates based on their modeled photo-eccentric effect \citep{dawson2012}. TKS will also explore the architectures of systems with multiple planets (in particular within the habitable zone) by searching for signals induced by additional planets with non-transiting geometries or those that reside on longer orbital periods.

\subsubsection{Spectroscopy of Exoplanet Atmospheres} \label{sec:SC3}

Multi-wavelength observations during transit and eclipse can constrain planet properties such as the admixtures of constituent gasses, the presence of clouds and/or hazes, and the rate of atmospheric loss. For sub-Neptune-size planets, atmospheric observations are particularly valuable as they can disambiguate between degenerate models of interior bulk composition \citep{rogers2010}. In turn, derived properties from these observations may probe the location within the protoplanetary disk where the planet formed, the interaction between its atmosphere and the stellar radiation field, and questions in planetary habitability \citep{mordasini2016}. Current measurements of sub-Jovian atmospheres have been sample-size-limited due to the scarcity of such targets around bright stars \citep{crossfield2017}.

In fact, with the launch of the \emph{James Webb Space Telescope} on the horizon, precision mass measurements of small TESS planets amenable to atmospheric characterization are critical for planning future follow-up observations. Exoplanets require precise ($\geq$5$\sigma$) mass measurements to break the degeneracy between surface gravity and atmospheric mean molecular weight when interpreting their transmission spectra \citep{batalha2019}. Therefore, TKS will follow up and measure $\geq$5$\sigma$ mass measurements in order to identify prime targets to be later observed with \jwst\ for atmospheric studies. 

\subsubsection{Evolved Host Stars} \label{sec:SC4}

Subgiant stars are frequently avoided in exoplanet surveys, with the exception of early RV surveys that aimed to probe the occurrence rates of gas-giant planets with stellar mass \citep{johnson2007,johnson2010,johnson2013}. For transiting planets, larger stars bias detections towards larger planets while for RV measurements for subgiants are noisier than their main-sequence counterparts \citep{luhn2019,luhn2020}. However, the relative time stars spend on the subgiant branch is comparatively small and therefore, a location on an HR-diagram conveniently provides a much more precise mass and age than that of a main-sequence star. Specifically in the \gaia\ era, an effective temperature, luminosity and metallicity alone are a powerful tool to characterize subgiants, including the planets that orbit them. 

Precise ages are valuable to place observational constraints on dynamical timescales for exoplanets. Previous studies have suggested that the dynamical timescales for processes like circularization or inward migration through tidal dissipation are strongly dependent on the scaled semi-major axis of the system \citep[\ator, see][]{zahn1977,hut1981,zahn1989}. Since this property is most rapidly changing for subgiant stars, stellar evolution is expected to affect tidal circularization timescales of close-in gas-giant planets, producing a transient population of mildly eccentric planets orbiting evolved stars \citep{villaver2014}. \kepler\ and K2 data have yielded intriguing evidence supporting this theory, but were based on a small sample of planets \citep{grunblatt2018,chontos2019}. By building up a more statistically-significant population of planets orbiting subgiants, TKS will investigate the role of stellar evolution in shaping post-main-sequence planetary systems.

Another advantage to studying planets orbiting evolved stars is that the likelihood of performing asteroseismology, since the amplitude of the oscillations increases with stellar luminosity. Ensemble studies of exoplanets orbiting asteroseismic stars have provided some of the most precise planet properties to date, revealing features like the hot sub-Neptune desert \citep{lundkvist2016} and the radius valley \citep{vaneylen2018}. First examples for asteroseismology-exoplanet synergies with TESS include transiting planets around TOI 197 \citep{huber2019} and TOI 257 \citep{addison2021}, as well as asteroseismic detections in solar analogs that are prime targets for future direct imaging missions \citep{chontos2020}.

\subsection{Technical Outcomes}

\subsubsection{Host Star Characterization} \label{sec:toa}

Since transiting exoplanets are discovered indirectly, fundamental properties of planet hosts are essential to realize the full scientific return of TESS. High-resolution spectroscopy allows the precise determination of effective temperatures and chemical abundances, which combined with Gaia parallaxes allows the precise characterization of exoplanet host star properties such as stellar masses \citep{berger2018,berger2020}. The results of the homogeneous stellar characterization of TESS host stars using Keck/HIRES, including TESS planet hosts beyond the target sample described here, will be presented in a future study.

\begin{figure*}
\centering
\includegraphics[width=\linewidth]{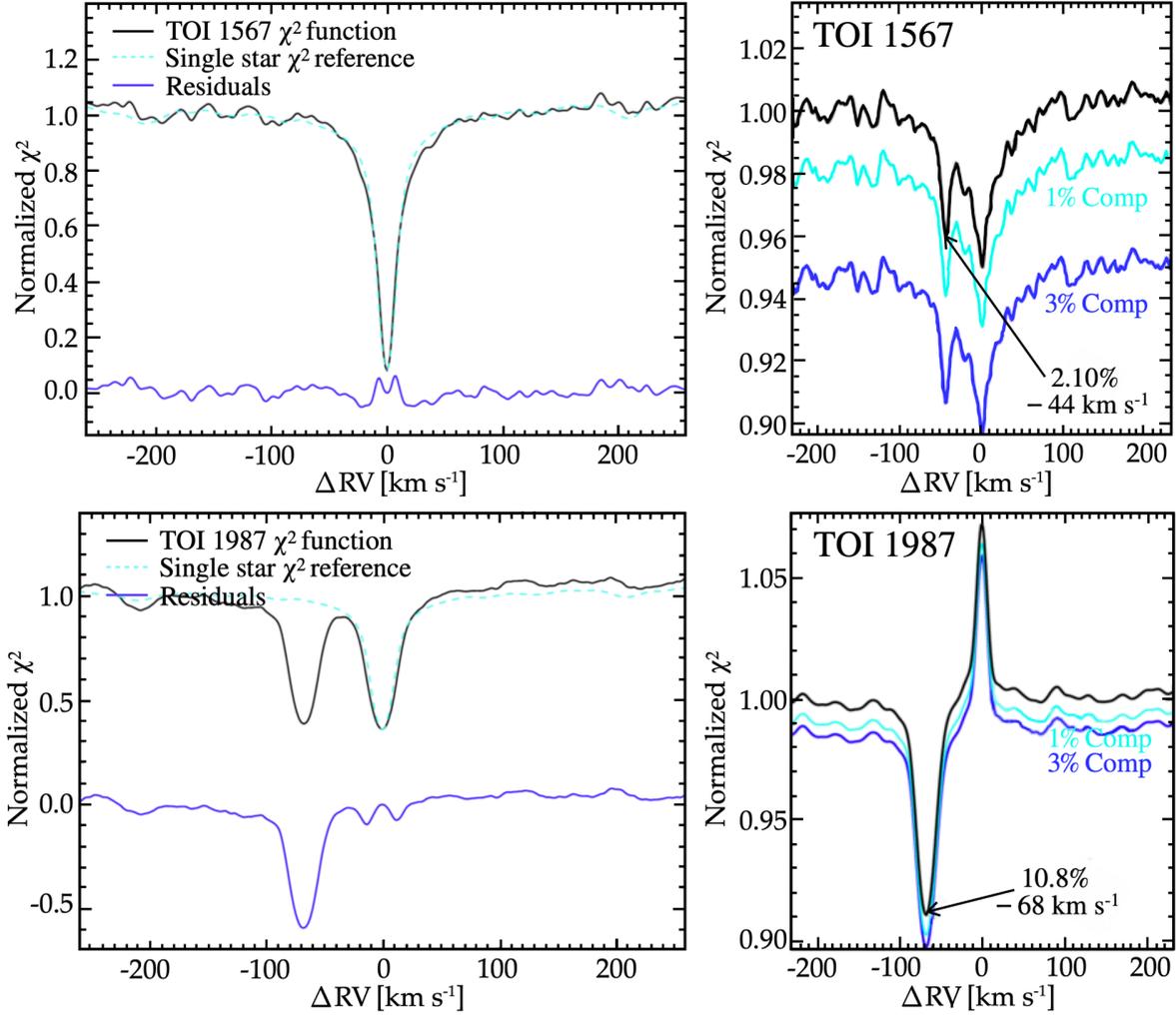}
\caption{SB2 vetting plots for TOI 1567 and TOI 1987 from \texttt{ReaMatch} \citep{kolbl2015}. The left panels show the $\chi^2$ value as a function of Doppler shift ($\Delta RV$) for the target of interest. Any deviations from the characteristic single star reference suggest a stellar companion and is usually sufficient for detecting bright companions (bottom). Any asymmetries or offsets from $\Delta RV=0$ in the right panels are typically indicative of a faint companion.}
\label{fig:FPs}
\end{figure*}

\begin{figure*}
\centering
\includegraphics[width=\linewidth]{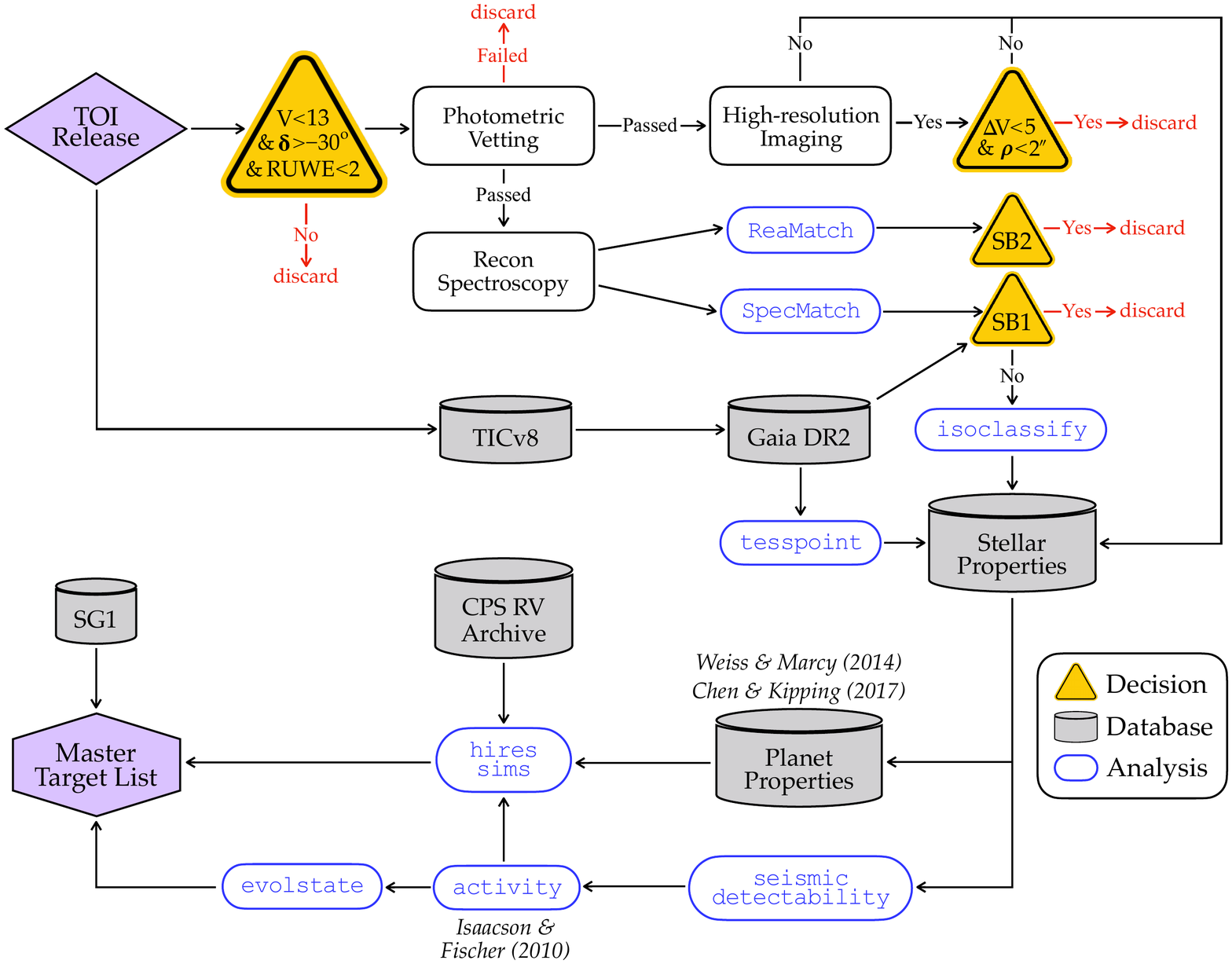}
\caption{Flowchart summary of the TKS target vetting (Section \ref{sec:vet}) and master target list construction (Section \ref{sec:master}) processes. Software packages are plotted in blue, while databases are shown in gray. All cuts and/or decision tree steps are in yellow and red represents points where targets were removed from the TKS selection pool.}
\label{fig:flowvet}
\end{figure*}

\subsubsection{Understanding Stellar Doppler Noise} \label{sec:tob}

Various stellar phenomena such as spot modulation, granulation, and chromospheric activity, produce apparent RV shifts or ``jitter'' that compete with planetary signals and thus, make more active stars unfavorable targets for precise radial velocity (PRV) work. Fortunately, TESS enables a link between photospheric variability and observed RV shifts by providing precise photometry for stars which already have long-term ground-based RV monitoring. In addition, activity studies benefit immensely when RVs and photometry are collected contemporaneously. For such datasets, jitter may be mitigated to reveal exoplanet signals that would otherwise be inaccessible \citep{lopezmorales2016}. 

The TESS-Keck Survey will leverage TESS to better understand the astrophysical processes causing jitter and have developed techniques to mitigate and remove it. Our survey design incorporates jitter tests for all included stars before designating particular targets of specific interest to this science case.

\section{Target Vetting} \label{sec:vet}

\subsection{Photometry} \label{sec:PV}

All TKS targets underwent a series of vetting procedures before being selected for precision RV follow-up. Targets are drawn from the database of TESS Objects of Interest \citep[TOI;][]{guerrero2021}, which are made publicly available\footnote{\url{https://tev.mit.edu/}}. Targets below a declination of $-30^{o}$ were excluded based on Keck observability and resources. A visual magnitude cut ($V<13$) was implemented where available, otherwise the TESS magnitude was used ($T<13$). Targets with a higher Gaia re-normalized unit weight error\footnote{\url{https://gea.esac.esa.int/archive/}} (RUWE $>2$), indicative of an unresolved companion, were also excluded from target selection \citep{belokurov2020,evans2018b}. Additional cuts were implemented for candidate hosts with \teff\ $>$ 6500 K and planet candidates with \radiusp\ $>$ 22 \radiuse.

All remaining targets were individually vetted by a member of the TKS photometric ``tiger'' team. Data validation (DV) products from the SPOC \citep[Science Processing Operations Center;][]{jenkins2016} and the QLP \citep[Quick Look Pipeline;][]{huang2020a,huang2020b} pipelines were inspected, ruling out possible false positive scenarios. Close analysis was given to metrics such as the ghost diagnostic and odd-even transit differences. Additionally, TOIs with signal-to-noise (\snr) less than 10 were excluded, since the planet radius uncertainty would dominate the uncertainty in the bulk density for such targets. 

A special set of selection rules were implemented for out-of-transit centroid offsets, where targets with a $<3\sigma$ offset would always pass this vetting step. For centroid offsets between $3\sigma$ and $5\sigma$, a target would still pass if the offset was $<21''$, unless the centroid offset was at a nearby star according to the DV report. For $>5\sigma$ offsets, the target would still pass if $T<7.5$, since the out-of-transit centroid offsets are unreliable if a target is saturated. 

Finally, a search for close companions was performed for the remaining targets using their Gaia DR2 coordinates and the MAST (Mikulski Archive for Space Telescopes) Portal\footnote{\url{https://mast.stsci.edu/portal/Mashup/Clients/Mast/Portal.html}}. Any targets with bright ($\Delta V<5$), close (separation $< 2"$) companions were not considered for further follow-up. Additionally, any star with a $<1"$ companion was removed irregardless of its magnitude since it would be challenging to resolve in poorer weather conditions and could possibly contaminate flux on the slit during observations.

\subsection{Spectroscopy} \label{sec:SV}

Once a TOI passed all photometric vetting steps, the target was queued for a reconnaissance (or recon) spectrum. A recon spectrum is an iodine-free exposure with an \snr\ $\approx40/\mathrm{pixel}$ to check for rapid rotation and spectroscopic false positives. Recon spectra are processed using \texttt{SpecMatch} \citep{petigura2015} and \texttt{ReaMatch} \citep{kolbl2015}.

\texttt{SpecMatch} estimates reliable spectroscopic stellar parameters \{\feh, \teff, \vsini, \logg\} from spectra with \snr\ $\lesssim$ $40/\mathrm{pixel}$ \citep{petigura2015}. We used results from \texttt{SpecMatch-Synth}, which uses synthetic model atmopheres for stars with effective temperatures $4700-6500 \mathrm{K}$. \texttt{SpecMatch-Emp} \citep{yee2017}, which relies on a comparison with empirical templates, was used for late-type stars (\teff\ $\rm \lesssim 4700\,K$). Smaller planet candidates orbiting stars with elevated \vsini\ ($\geq$ 8 \kms) were not typically considered due to the increased number of observations required to achieve a similar mass precision (see \S \ref{sec:noise} for more details). 

\begin{table}[]
\centering
\caption{Spectroscopic false positives or high \vsini\ targets not suitable for PRV work identified through TKS, as described in Section \ref{sec:SV}.}
\begin{tabular}{llcl}
\hline\hline
TIC ID & TOI & Disposition$^{*}$ & Note\\
\hline
405904232 & 1312 & SB1 & \\
129539786 & 1367 & -- & \vsini\ \equal\ 12 \kms \\
148782377 & 1415 & -- & \vsini\ \equal\ 17 \kms \\
411608801 & 1494 & -- & \vsini\ \equal\ 15 \kms \\
376637093 & 1516 & -- & \vsini\ \equal\ 10 \kms \\
259151170 & 1567 & SB2 & \\
285677945 & 1571 & -- & \vsini\ \equal\ 20 \kms \\
138017750 & 1608 & -- & \vsini\ \equal\ 44 \kms \\ 
184679932 & 1645 & SB1 & \vsini\ \equal\ 30 \kms \\
468828873 & 1672 & SB2 &  \\
58542531 & 1683 & SB2 & \\
103448870 & 1687 & -- & \vsini\ \equal\ 15 \kms \\
461662295 & 1711 & SB1 & \\
356978132 & 1755 & -- & \vsini\ \equal\ 11 \kms \\
450327768 & 1788 & -- & \vsini\ \equal\ 12 \kms  \\
330799746 & 1818 & SB1 & \vsini\ \equal\ 10 \kms \\
20182165 & 1830 & -- & \vsini\ \equal\ 29 \kms \\
349088467 & 1987 & SB2 & \vsini\ \equal\ 110 \kms \\
\hline
\multicolumn{4}{l}{{\sc \textbf{Note --}}} \\
\multicolumn{4}{l}{$^{*}$ Dispositions in this table represent the false positive classification} \\
\multicolumn{4}{l}{suggested by initial TKS reconnaissance observations and therefore} \\
\multicolumn{4}{l}{do not necessarily reflect the most up-to-date TFOP dispositions.} \\
\end{tabular}
\label{tab:FPs}
\end{table}

The \texttt{ReaMatch} algorithm \citep{kolbl2015} is a spectroscopic method used to identify faint stellar companions in double-lined spectroscopic binaries (SB2). Figure \ref{fig:FPs} shows the SB2 vetting results from \texttt{ReaMatch} for TOIs 1567 and 1987. The left panels of Figure \ref{fig:FPs} show a cross-correlation of the observed, rest-frame corrected spectrum with an NSO Solar spectrum. Significant deviations to a single star function in the residuals are typically sufficient to identify secondary stars which contribute a significant fraction of flux, as for TOI-1987.

To check for possible faint secondary lines, a best-fit spectrum is matched from the \texttt{SpecMatch} library, which is then broadened and diluted to match the absorption lines of the primary. The matched spectrum is subtracted from the observed spectrum, the residuals are renormalized and analyzed in a similar way to the primary. The right panels of Figure \ref{fig:FPs} show the residual CCFs, where any large deviation from $\Delta RV \sim$ $0$ \kms\ is evidence for a second set of absorption lines. However, for stellar companions with similar radial velocities to the primary, spectral lines can be blended with the primary and therefore \texttt{ReaMatch} is limited to Doppler shifts $> 10$ \kms. 

Finally, systemic radial velocities were computed according to the methodology of \citet{chubak2012} and compared to Gaia to identify single-lined spectroscopic binary (SB1). Targets for which the systemic and Gaia RV disagreed by more than 5 \kms\ typically ``failed'' this vetting step. However, for borderline cases (i.e. where systemic and Gaia RVs differed between 5 and 10 \kms\ or other ambiguous false positive detections), a second recon spectrum was taken to test for significant linear trends that provided additional evidence of the SB1-like nature.

Targets that failed any of these analyses were reported to the TESS Follow-up Observing Program\footnote{\url{https://tess.mit.edu/followup/}} (TFOP) Recon Spectroscopy Working Group (WG) SG2. In addition, stellar properties derived from HIRES spectra through the TKS project were/are made available to the community via the Exoplanet Follow-up Observing Program for TESS (ExoFOP-TESS) website\footnote{\url{https://exofop.ipac.caltech.edu/tess/}}. Reported stellar parameters include spectroscopic parameters \teff, \logg, \feh, and \vsini, as well as stellar mass and radius derived using the spectroscopic parameters as input to \texttt{isoclassify} \citep{huber2017,berger2018}. A list of spectroscopic false positives and/or stars with rapid rotation is provided in Table \ref{tab:FPs}.

\begin{figure*}
\centering
\includegraphics[width=\linewidth]{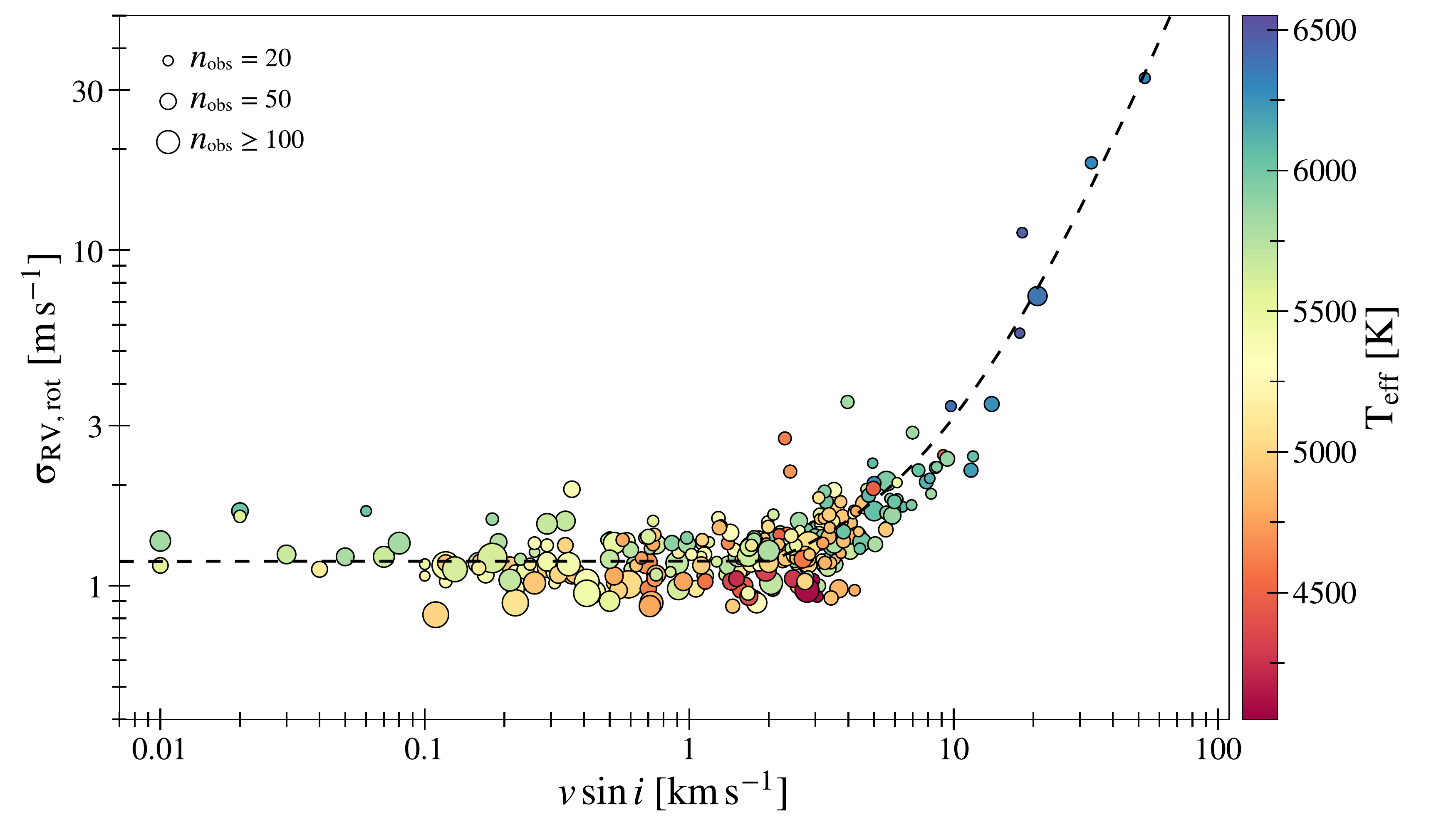}
\caption{Median RV uncertainty versus \vsini\ for all stars $V<10$ with at least 20 HIRES observations and effective temperatures between $4000-6500 \,\,\mathrm{K}$. Markers are sized by the number of HIRES observations and colored by the effective temperature of the star. The black dashed line represents our empirical fit to the data, which is defined by the polynomial fit in Section \ref{sec:noise} (Equation \ref{eqn:rot}).}
\label{fig:vsini}
\end{figure*}

\subsection{High Resolution Imaging}


Multiplicity has historically been disfavored during follow-up target selection processes since nearby companions can contaminate the spectrum through the slit, resulting in less precise RVs. Similar to the \kepler\ efforts, the TFOP High-Resolution Imaging WG (hereafter referred to as SG3) has been reporting any stellar companions to ExoFOP TESS \citep[e.g.,][]{ziegler2020}. For TOIs that had adaptive optics, speckle or lucky imaging information available, we excluded TOIs for which the data or SG3 comments indicated the presence of a bright ($\Delta V < 5$ or $\Delta K < 5$) and nearby (separation $<2$'') contaminant. For targets that had not yet been followed up by SG3, this was not a requirement for the target selection process. 

There are two TOIs that failed this vetting step but made it to our final TKS target sample: TOI-1288, which only marginally failed by the presence of two faint companions, and TOI-1443, which was added to an individual science case that did not preferentially select against binaries.

\section{Master Target List} \label{sec:master}

\subsection{TOI Information}

In addition to the target vetting, we constructed a master target list that provided additional information for TESS targets. This is shown in Figure \ref{fig:flowvet}, which also starts from a TOI release but runs in parallel to the target vetting and therefore, was performed for all TOIs.

For each TOI, the TESS Input Catalogue \citep[version 8, TICv8;][]{stassun2019} was queried using to fetch stellar properties (e.g., radius, etc.), available photometry (i.e. Johnson $V$ and 2-MASS $JHK$) and the associated Gaia Data Release 2 \citep[DR2;][]{gaia2016,evans2018a,gaia2018} source ID. The Gaia ID was then used to collect astrometric and photometric Gaia parameters. The TICv8 position was used with  \texttt{tesspoint} \citep{tesspoint} to determine how many sectors a given TESS target would be observed for. This was important for the TKS evolved stars (SC4) and Doppler noise (TB) aspects, both which benefit from longer baselines. The evolutionary state of the star was calculated using  \texttt{evolstate}\footnote{\url{https://github.com/danxhuber/evolstate/}}, given a star's effective temperature and radius  \citep{huber2017,berger2018}. For all steps that involved calculations using stellar properties, information first came from SpecMatch when a HIRES spectrum was available (see Section 3.2 for more details), otherwise the TICv8 was used. In the event that neither were available, stellar properties provided in the original TOI table were used.

Using the TOI planet radii, masses were computed using the following relations:

\small
\begin{equation*}
\mathrm{M_{p}} = \left\{
\begin{array}{ll}
\rho_{\mathrm{p}} = 2.43+3.39 \Big(\frac{\mathrm{R_p}}{\mathrm{R_{\oplus}}}\Big) \, \mathrm{g\,cm}^3 & \mathrm{R_p} < 1.5 \,\mathrm{R_{\oplus}}\\
 2.69 \Big(\frac{\mathrm{R_p}}{\mathrm{R_{\oplus}}}\Big) & 1.5\, \mathrm{R_{\oplus}} \leq \mathrm{R_p} < 4\, \mathrm{R_{\oplus}}\\
1.24 \Big(\frac{\mathrm{R_p}}{\mathrm{R_{\oplus}}}\Big)^{1.7} &  4\, \mathrm{R_{\oplus}} \leq \mathrm{R_p} < 11.3\, \mathrm{R_{\oplus}}\\
317.83  &  \mathrm{R_p} \geq 11.3\, \mathrm{R_{\oplus}}\\
\end{array}
\right.
\end{equation*}
\normalsize
The relations are based on \citet{weiss2014} for small planets, \citet{chen2017} for intermediate-mass planets, and assuming a Jupiter mass for planets larger than Jupiter. Expected Doppler amplitudes were calculated using

\begin{equation}
K_{\star,\mathrm{exp}} = \bigg(\frac{2 \pi G}{P}\bigg)^{1/3} \frac{\mathrm{M_p}}{(M_p + M_{\star})^{2/3}},
\end{equation}
where P is the orbital period, $G$ is the gravitational constant, and M$_{p}$ and M$_{\star}$ are the masses of the planet and star, assuming the orbit is aligned ($i=90^{o}$) and circular ($e=0$).

The CPS RV archive\footnote{\url{https://jump.caltech.edu/}} was queried for targets that had existing spectra, which was important for estimating astrophysical ``jitter'' (Section \ref{sec:noise}) and expected exposure times (Section \ref{sec:hires}). As a final step, TOI dispositions were updated with the TESS Follow-up Observing Program (TFOP) Working Group\footnote{\url{https://tess.mit.edu/followup/}} (WG) Sub Group 1 (hereafter referred to as SG1). TOIs with unfavorable or ambiguous dispositions (e.g. APC, BEB, FA, FP, NEB, etc.)\footnote{Ambiguous planet candidate (APC), blended eclipsing binary (BEB), false alarm (FA), false positive (FP), nearby eclipsing binary (NEB), etc.} were removed.

\subsection{Astrophysical Noise} \label{sec:noise}
 
To account for targets that would be challenging for PRV work, we estimated a lower limit for a single RV uncertainty by considering the contributions from various stellar sources. Specifically, we include effects from stellar rotation and magnetic activity, as well as effects due to near-surface processes like convection, granulation, and acoustic oscillations. 

To estimate the Doppler noise contribution due to stellar activity (\sigact), we used the empirical relation from \citet{isaacson2010} for stars that had the Ca \rom{2} S-index and color ($B-V$) information available. For granulation and p-mode oscillations, $\sigma_{\mathrm{gran}}$ was estimated using the effective temperature and surface gravity as inputs to the RV jitter prediction code\footnote{\url{https://github.com/Jieyu126/Jitter/}} by \citet{yu2018}.

To investigate how rotation affected a typical RV observation, we calculated the median RV uncertainty for all targets from the CPS RV archive. We kept only stars with effective temperatures between $4000-6500 \,\,\mathrm{K}$ with at least 20 HIRES observations which are brighter than $V<10$ to to remove targets with observations that could be dominated by photon noise. Figure \ref{fig:vsini} shows the median HIRES RV uncertainty as a function of \vsini. For \vsini\ $<2$ \kms, the effects of rotational broadening are negligible and therefore the RV precision is set by the typical noise floor of the instrument and is 1.18 \ms\ for HIRES. For moderate and  high \vsini\ ($\geq 2$ \kms), we fit a second order polynomial to estimate \sigrot,

\begin{equation} \label{eqn:rot}
\sigma_{\mathrm{rot}} = 0.867 + 0.140\,(v\sin i) + 0.009\,(v\sin i)^{2},
\end{equation}

\noindent using a least-squares minimization, where \sigrot\ and \vsini\ are in units of \ms\ and \kms, respectively.

\subsection{HIRES Observing Simulations} \label{sec:hires}

Exposure times were scaled based on the canonical exposure time of 100 seconds for an iodine-in observation on a $V=8$ star in nominal conditions. For iodine-out exposures, the HIRES throughput is $\sim 30\%$ higher and was therefore factored in when relevant (e.g., recon spectra, templates). The required counts, which determines the photon-limited RV precision (\sigrv), varied between science cases. A maximum exposure time of 30 minutes was implemented for all TKS observations.

The total single measurement uncertainty was then calculated using

\begin{equation}\label{eqn:sigma}
\sigma_{\mathrm{tot}}^2 = \sigma_{\mathrm{RV}}^2 + \sigma_{\mathrm{act}}^2 + \sigma_{\mathrm{gran}}^2 + \sigma_{\mathrm{rot}}^2,
\end{equation}
which is the contribution from individual noise sources added in quadrature. The measurement uncertainty scales as \sigtot $\propto$ \nobs$^{-1/2}$, where the default number of observations for most TKS science cases was $\rm N_{obs}=60$. Some exceptions that warranted more observations included the multis (SC2C) and activity (TB) science cases, which instead required 100 observations. On the other hand, the distant giants (SC2A) science case preferred less precision and observations ($\rm N_{obs}=15$) in exchange for more targets and longer baselines. 

An average overhead of $120\,\mathrm{s}$ was charged per target per observation that accounted for telescope slew time and CCD readout times. For science cases that depended on a certain mass precision (e.g., SC3), total nominal exposure times to achieve 5$\sigma$ (\tfive) and 10$\sigma$ (\tten) masses were calculated for all TOIs. Finally, to estimate realistic target costs, calculations also included archival HIRES data and factored in if high-resolution templates were already available.

\begin{figure*}
\centering
\includegraphics[width=\linewidth]{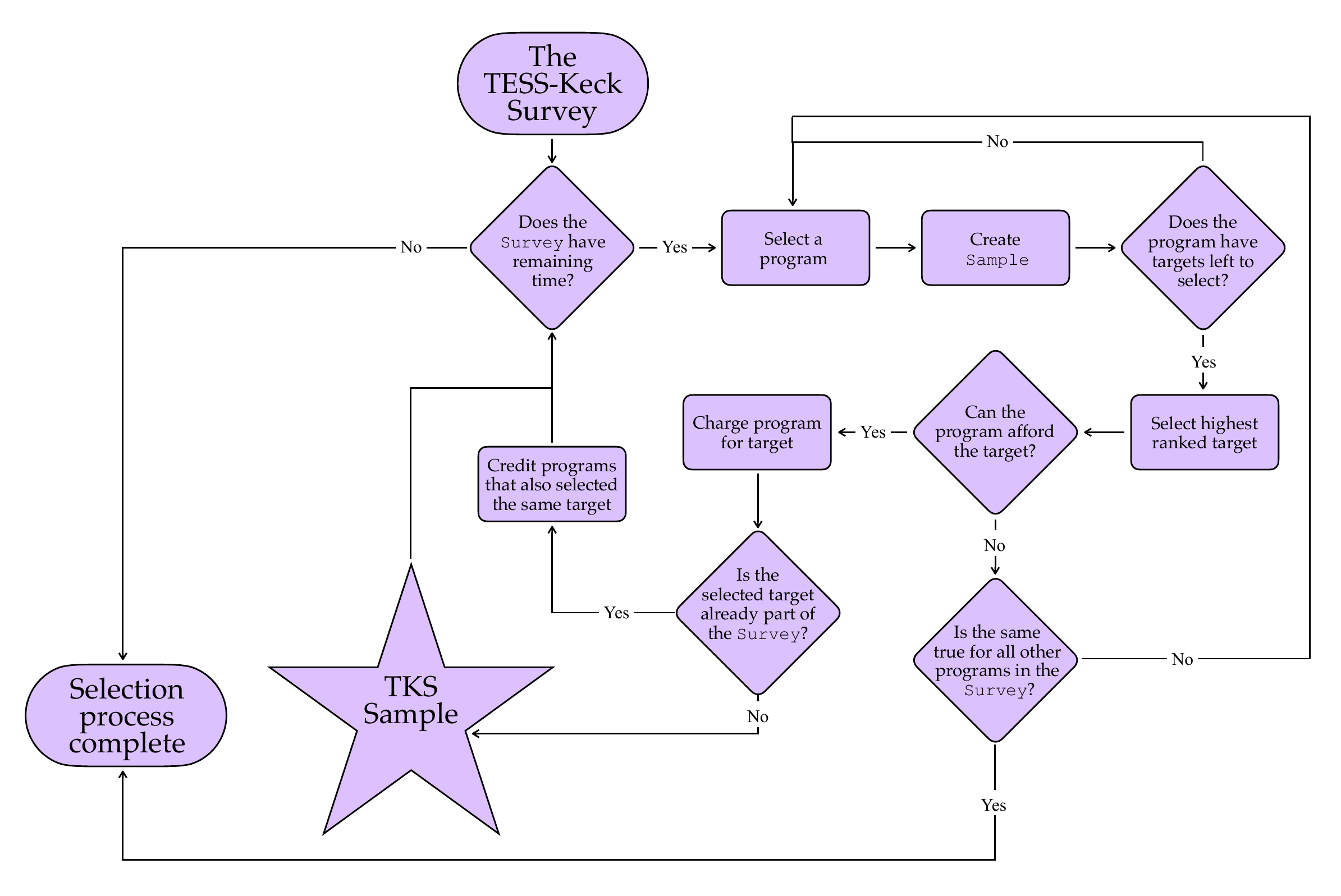}
\caption{Flow diagram for the survey target prioritization algorithm. The \texttt{Survey} class contains all programmatic survey information as well as the vetted sample. The selection process will continue until either all of the time in the Survey is used or all science cases in the \texttt{Survey} are no longer able to make selections.}
\label{fig:flowselect}
\end{figure*}

\subsection{Individual Science Case Selection Criteria} \label{sec:scc}


\subsubsection*{\textsc{SC1A: P\lowercase{lanet} R\lowercase{adius} G\lowercase{ap}}}

The focus of SC1A is bulk properties of planets in the radius valley. However, to increase overlap with other science cases, the radius limits were expanded to include any planets in the range, 1 \radiuse\ $<$ \radiusp\ $<$ 3.5 \radiuse.

\subsubsection*{\textsc{SC1B: S\lowercase{tellar} F\lowercase{lux} \& G\lowercase{aseous} E\lowercase{nvelopes}}}

SC1B focuses on a narrow range of planet sizes that are exposed to very different stellar environments. In particular, SC1B focused on smaller planets with sizes, 1 \radiuse\ $<$ \radiusp\ $<$ 4 \radiuse, with orbital periods up to 100 days. To avoid sample biases, targets in this parameter space were selected as uniformly and as randomly as possible. To achieve this, 12 bins were created in roughly equal bin sizes in 2-dimensional log-linear space, with planet radius bin edges of \radiusp\ $\sim$[1, 2, 3, 4] \radiuse\ (i.e. linear-space) and orbital period bin edges of P $\sim$[1, 3, 10, 30, 100] days (i.e. log-space).

\subsubsection*{\textsc{SC1C: U\lowercase{ltra}-S\lowercase{hort} P\lowercase{eriod} P\lowercase{lanets}}}

SC1C is primarily interested in differentiating between the smaller ($<$ 2 \radiuse) USP population that appear to have $\sim$Earth-like compositions and the recently discovered sub-day hot Neptunes \citep[e.g., TOI$-$849 b;][]{armstrong2020}. Therefore, the selection criteria for SC1C includes both classes of sub-Jovian planets, with \radiusp\ $<8$ \radiuse\ and \sinc\ $>650$ \since.

\subsubsection*{\textsc{SC1D: H\lowercase{abitable} Z\lowercase{one} P\lowercase{lanets}}}

We selected potential habitable zone planets based on their incident flux. We adopted a conservative limit of $<10$ \since, and required all SC1D targets to orbit main-sequence stars.



\subsubsection*{\textsc{SC1E: P\lowercase{lanet}-S\lowercase{tar} C\lowercase{orrelations}}}

To enable statistical investigations of possible correlations between stellar and planetary properties, we selected planets orbiting stars with a wide range of masses and metallicities. In order to ensure that low-mass stars were observed despite their faintness, TOI-1467 and TOI-1801 were added manually as high-priority targets (see Section \ref{sec:sample}). Due to the broad parameter space covered by this science case, almost all selected targets had overlap with multiple science cases.

\subsubsection*{\textsc{SC2A: D\lowercase{istant} G\lowercase{iants}}}

We selected all main-sequence stars brighter than $V=12$ that have an approximately solar-like mass ($0.5$\msun\ $\leq$ \mstar\ $\leq 1.5$\msun). We avoided rapidly rotating and/or active stars by also applying cuts in \teff\ ($\leq 6200$ K $\approx$ Kraft Break) and \logrhk\ ($\leq -4.7$). For stars that had recon spectra available and thus a measured projected velocity, we excluded anything with a \vsini\ $\geq 5$ \kms. We also excluded stars with Gaia RUWE $>1.3$.

\begin{figure*}
\centering
\includegraphics[width=\linewidth]{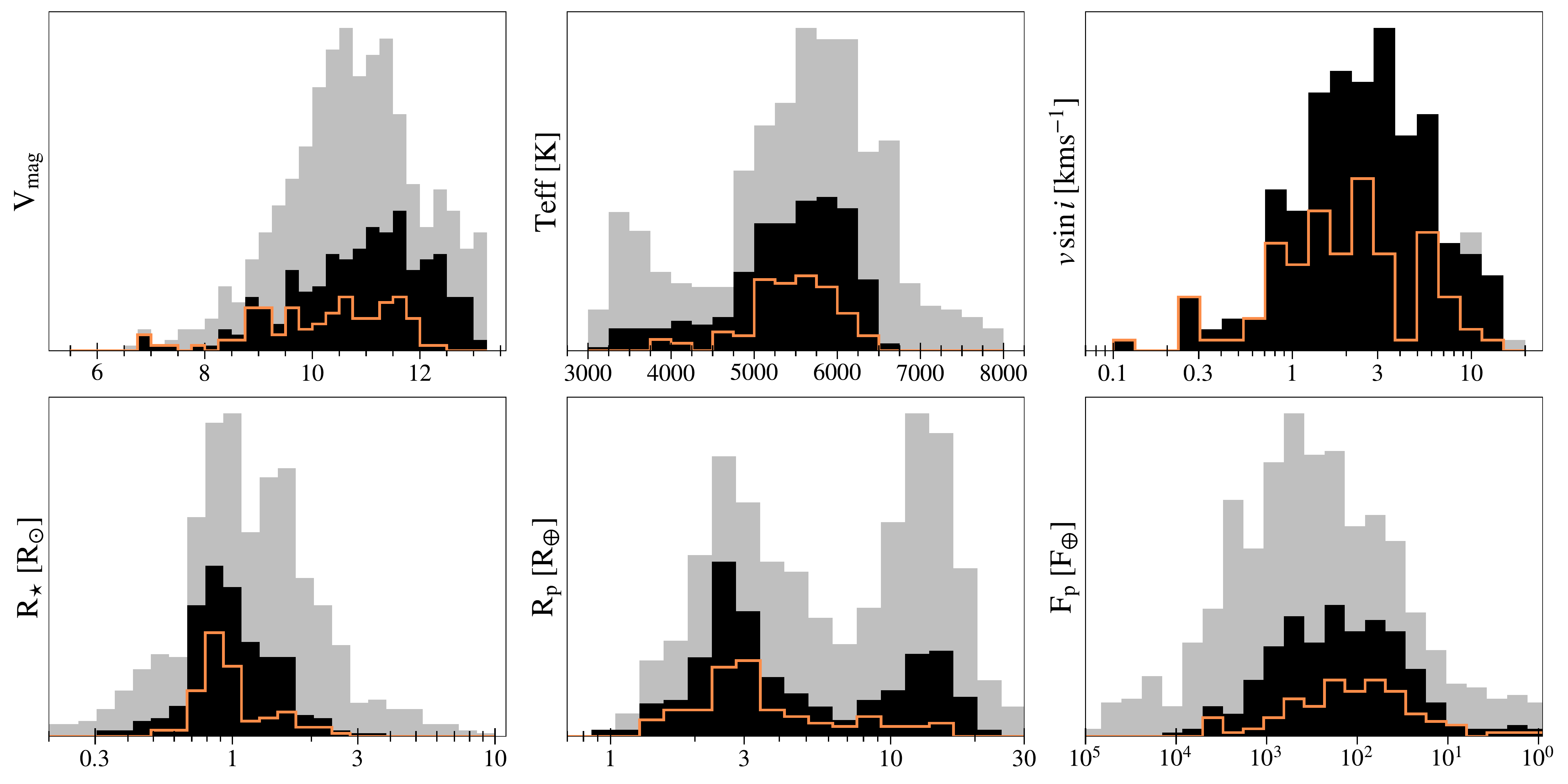}
\caption{All TESS Objects of Interest above a declination of $\delta >= -30^{o}$ with a low Gaia RUWE $<$ 2 (gray sample). The black sample comprises all TOIs that passed every vetting step and therefore available for target selection. The final TKS sample that was selected by the target prioritization algorithm is shown by the \plotcolor\ line.}
\label{fig:sample}
\end{figure*}

We only selected targets with a multiple event statistics (MES) of at least 12, which indicated high-quality transit detections based on a visual inspection of SPOC DV reports. Only systems that contained at least one transiting planet smaller than 10 \radiuse\ were kept (Van Zandt et al., in prep).

\subsubsection*{\textsc{SC2B: O\lowercase{bliquities} \lowercase{and} E\lowercase{ccentricities}}}

Obliquity targets were selected on a case-by-case basis. To select high-probability eccentric planet candidates, we performed transit fitting to measure the photo-eccentric effect, which compares the observed transit duration with the expected duration for a circular orbit \citep{dawson2012,kipping2012}. Transit duration is related to the eccentricity (e) and the argument of periastron ($\omega$) by
\begin{equation}
\label{eqn:duration}
T_{14} = \left(\frac{R_* P}{\pi a}\sqrt{1-b^2}\right)\frac{\sqrt{1-e^2}}{1+e\sin{\omega}},
\end{equation}
\noindent where $T_{14}$ is the total transit duration (i.e. from first to final contact), \rstar\ is the stellar radius, $P$ is the orbital period, $b$ is the impact parameter, and $a$ is the planet's semi-major axis \citep{winn2010transits}. 

Instead of using the standard transit parameter $a/R_*$, we reparametrized our model in terms of stellar density ($\rho_*$), which maps to $a/R_*$ through Kepler’s 3rd Law and can be measured independently. We combined this with Equation \ref{eqn:duration}, which yielded a new parameter ($\rho_{*,circ}$) that was directly sampled in our transit modeling using \texttt{exoplanet} \citep{fm2019}. We derived an eccentricity estimate by re-sampling from the posterior distribution of $\rho_{*,circ}$ and comparing these values to an independent stellar density measurement derived by combining HIRES spectroscopy, Gaia parallaxes, and isochrone models. We selected all targets for which zero eccentricity is ruled out at the 2-$\sigma$ level by this method (MacDougall et al., in prep).

\subsubsection*{\textsc{SC2C: M\lowercase{ultis}}} \label{sec:multis}

The only selection requirement for SC2C was the presence of more than one transiting planet. Systems with the highest planet multiplicity were ranked the highest. For systems with the same number of planet candidates, priority was given to lower cost targets (i.e. brighter and/or shared targets).

\subsubsection*{\textsc{SC3: A\lowercase{tmospheres}}} \label{sec:tsm}

The Transmission Spectroscopy Metric (TSM) from \citet{kempton2018} is an SNR proxy for atmospheric observations and was computed for all planet candidates. The TSM is defined as:

\begin{equation} \label{eqn:tsm}
\centering
\mathrm{TSM}_\mathrm{p} = S \times \frac{R_p^3   T_{eq}}{M_p  R_*^2} \times 10^{-0.2m_J},
\end{equation}
where $S$ is a dimensionless normalization constant,
\begin{equation*}
S = \left\{
\begin{array}{ll}
0.19, & R_p < 1.5\, R_\oplus \\
1.26, & 1.5 < R_p < 2.75\, R_\oplus \\
1.28, & 2.75 < R_p < 4.0\, R_\oplus \\
1.15, & 4.0 < R_p < 10\, R_\oplus.
\end{array}
\right.
\end{equation*}
Quantitatively, the TSM is the expected (or simulated) SNR from a 10-hour observation with \emph{JWST}-NIRISS, assuming a cloud-free, solar-metallicity, H$_2$-dominated atmosphere. For reference, a ``good" TSM for a sub-Neptune ($1.5 < R_p < 4.0$ R$_\oplus$) is roughly between 80 and 150 (see Table 1 in \citealt{kempton2018}).

\begin{figure*}
\centering
\includegraphics[width=\linewidth]{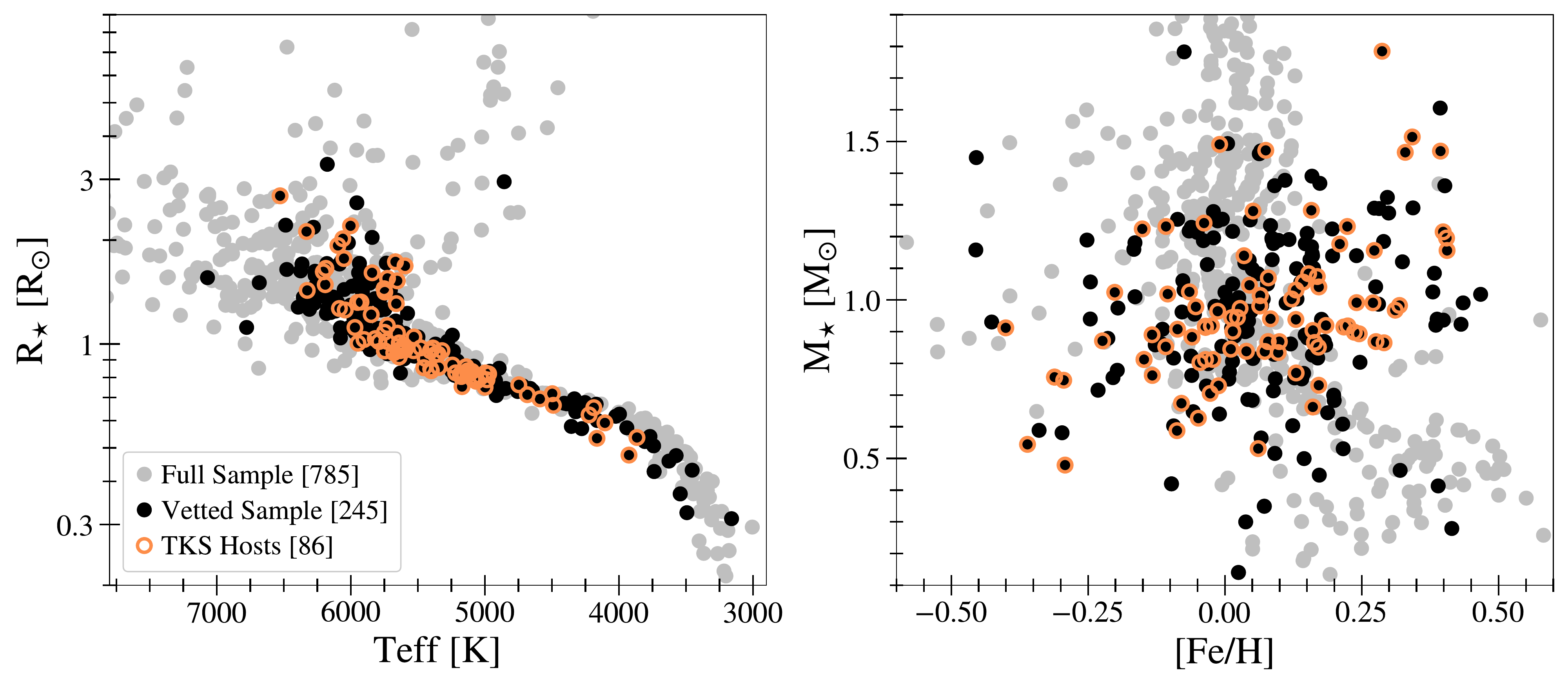}
\caption{TESS Objects of Interest plotted in an HR-diagram (left) and in stellar mass-metallicity space (right). Gray symbols represent all TOIs with $\delta >= -30^{o}$ and Gaia RUWE metric $<$ 2. Black symbols represent TOI hosts that passed all vetting steps and hence available during the target selection process, while the final TKS sample selected by the algorithm is shown in \plotcolor.}
\label{fig:hosts}
\end{figure*}

To rank targets, TOIs identified by the SPOC pipeline \citep{jenkins2016} were divided into log-uniform bins in planet radius (\radiusp), stellar effective temperature (\teff), and incident flux (\sinc). Bin edges were located at \radiusp\ = 1.0, 1.6, 2.5, 4.0, 6.3, and 11.2 $R_\oplus$, \teff\ = 2500, 3900, 5200, and 6500 K, and \sinc\ = 0.1, 1, 10, 100, 1e3, and 1e4 S$_\oplus$. Confirmed planets were added alongside the TESS candidates in order to identify TOIs which had potential to fill in sparsely-sampled regions of parameter space.

Ultimately, we divided the TSM by the expected Keck/HIRES exposure time required to obtain a 5$\sigma$ mass ($t_\mathrm{5\sigma,HIRES}$). The combined metric was used as a simple way to compromise between planets with the best atmospheric prospects and those orbiting bright hosts with reasonably-detectable (expected) Doppler amplitudes ($K_\mathrm{exp} \gtrsim 2$ m s$^{-1}$).

\subsubsection*{\textsc{SC4: E\lowercase{volved}}}

To identify higher priority targets that were likely to exhibit solar-like oscillations, asteroseismic detection probabilities were computed for all TOIs \citep{chaplin2011probs,schofield2019}. Broadly speaking, asteroseismic detection probabilities increase with luminosity, and longer time series. \citet{schofield2019} calculated probabilities for the TIC prior to the TESS launch, which required an assumption about the number of observed TESS sectors. Consequently, we recalculated detection probabilities using the actual number of observed TESS sectors. The selection criteria for SC4 required either an evolved star status (subgiant or red giant) for new planet candidate hosts or a predicted asteroseismic detection probability $\geq 0.5$.

\subsubsection*{\textsc{TB: A\lowercase{strophysical} D\lowercase{oppler} N\lowercase{oise} }}

For stars that had HIRES spectra available, an activity metric was computed for the Doppler noise aspect using traditional activity indicators for chromospheric emission like \logrhk\ and the estimated jitter. Targets with known rotation periods and longer baselines were typically given higher priorities. Gaussian weights centered on \logrhk $=-4.8$ ensured a moderate level of activity while downweighting both inactive stars, which would not be as scientifically interesting, as well as very active stars that would be challenging to characterize. Using the $\sigma_{\mathrm{tot}}$ metric calculated in Equation \ref{eqn:sigma} (Section \ref{sec:noise}), Gaussian weights were centered on a ``jitter'' value of 4 \ms\ using a similar logic as that for the chromospheric emission. Finally, brighter stars were ranked in ascending order, which was helpful for narrowing the targets down by an order of magnitude. However, more active targets require multiple considerations and therefore, the final high-priority high-cadence targets were hand-selected for this case. 

\begin{figure*}
\centering
\includegraphics[width=\linewidth]{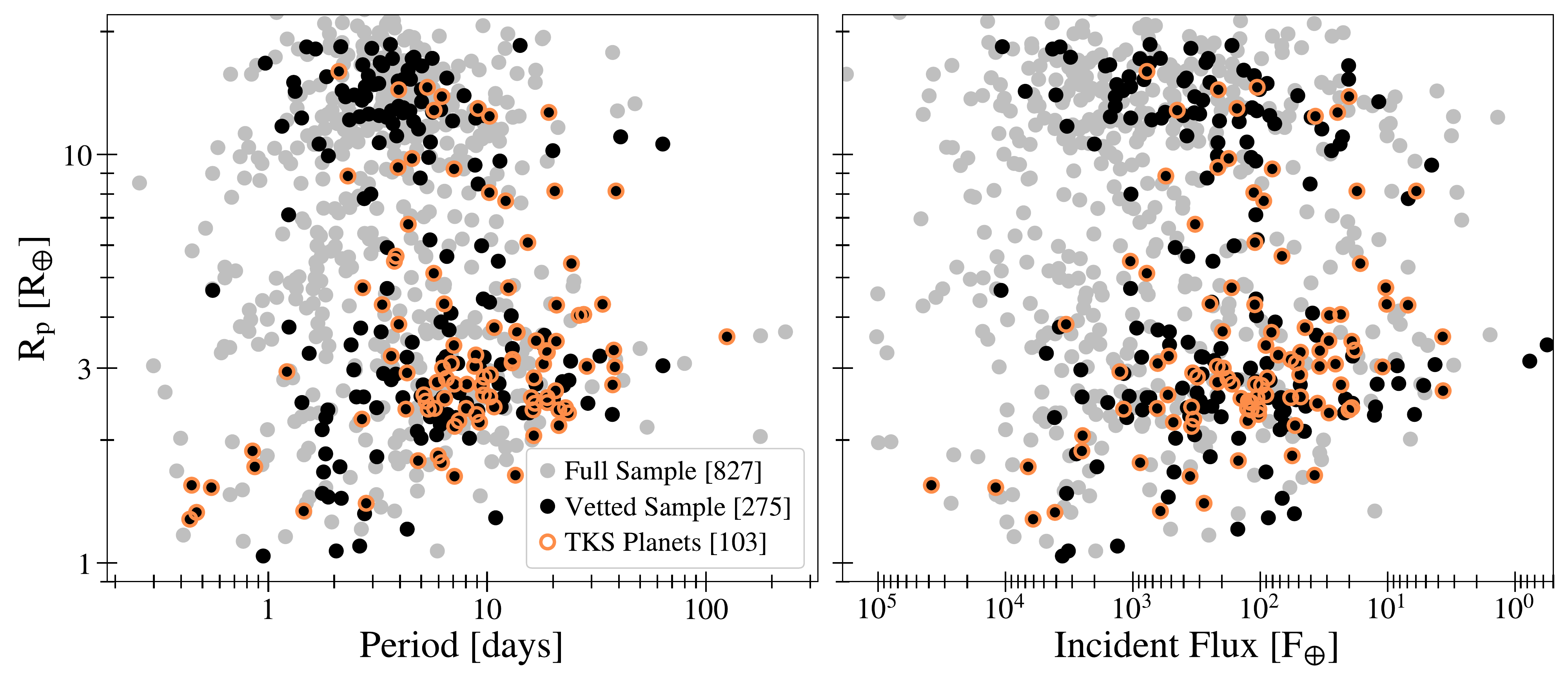}
\caption{Planet sizes of TOIs plotted versus period (left) and incident flux (right). Gray symbols represent all TOIs with declination $\delta >= -30^{o}$ and Gaia RUWE metric $<$ 2. Black symbols include all TOIs that passed every vetting step, as discussed in Section \ref{sec:vet}. The final selected TESS-Keck Survey sample is highlighted in \plotcolor, which includes \nhosts\ unique systems with a total of \nplanets\ planets.}
\label{fig:planets}
\end{figure*}

\section{Automated Target Selection}

\subsection{Motivation \& Initial Conditions}

Using the master target list described in Section \ref{sec:master}, the goal was to develop an automated and reproducible procedure to select targets given a set of science cases and time allocation. Specifically, a set of programs, $\rm P\supset\,$\{$\rm P_1,P_2,\dots,P_n$\} receive time allocations $\rm T\sim$\{$\rm T_1,T_2,\dots,T_n$\} of the total observing time allocation, $\rm T_{tot}$. A set of stars, S$\,\supset\,$\{$\rm S_1,S_2,\dots,S_m$\} have a set of observing times per star, t$\sim$\{$\rm t_1,t_2,\dots,t_m$\}, which can vary for each program. Each program has a set of ordinal rankings, R$\sim$\{$\rm R_1,R_2,\dots,R_m$\} for the set of stars S. In addition, some programs may want to update their rankings after each selection step and therefore,  individual program rankings have the flexibility to adapt and update with each iteration. This encourages cooperation because a program would rank a shared star more highly, which would ultimately cost that program (and other programs) less.

A key component of the target selection algorithm is the \texttt{Survey} class, which includes general survey information, science-case-specific requirements, and the vetted planet sample. Individual programs are able to specify an observing strategy (i.e. the total number of observations and required photon counts), any high priority or dropped targets, as well as the selection criteria to filter a program by. Upon the initialization of the \texttt{Survey} class, an initial accounting of the vetted sample occurs by determining the number of relevant science targets (or ``picks'') available for each program in the \texttt{Survey}. This is a required check to make sure that, at any point during the target selection process, the number of program picks does not exceed the available number of targets for a given science case. 

Each program within the \texttt{Survey} can specify how to prioritize and select their targets. Additionally, programs are not limited to a single criterion but instead can rank targets by providing a list of prioritization metrics. An example is SC2C, whose targets were first ranked by those with the highest planet multiplicity and then by target cost. For atmospheric targets, TOIs were prioritized by the TSM metric discussed in Section \ref{sec:tsm} and then sorted by cost. However, for situations that involved more complicated vetting (e.g., SC2A) or meticulous observing plans (e.g., SC2Bii) whose prioritization could not be automated, a list of rankings could be provided that did not change under any circumstances. This was particularly important for a majority of SC2 that focused on dynamical planets, which provided an outlet for target lists to ``override'' any automated process.

\subsection{Algorithm} \label{sec:algo}

\begin{figure*}
\centering
\includegraphics[width=\linewidth]{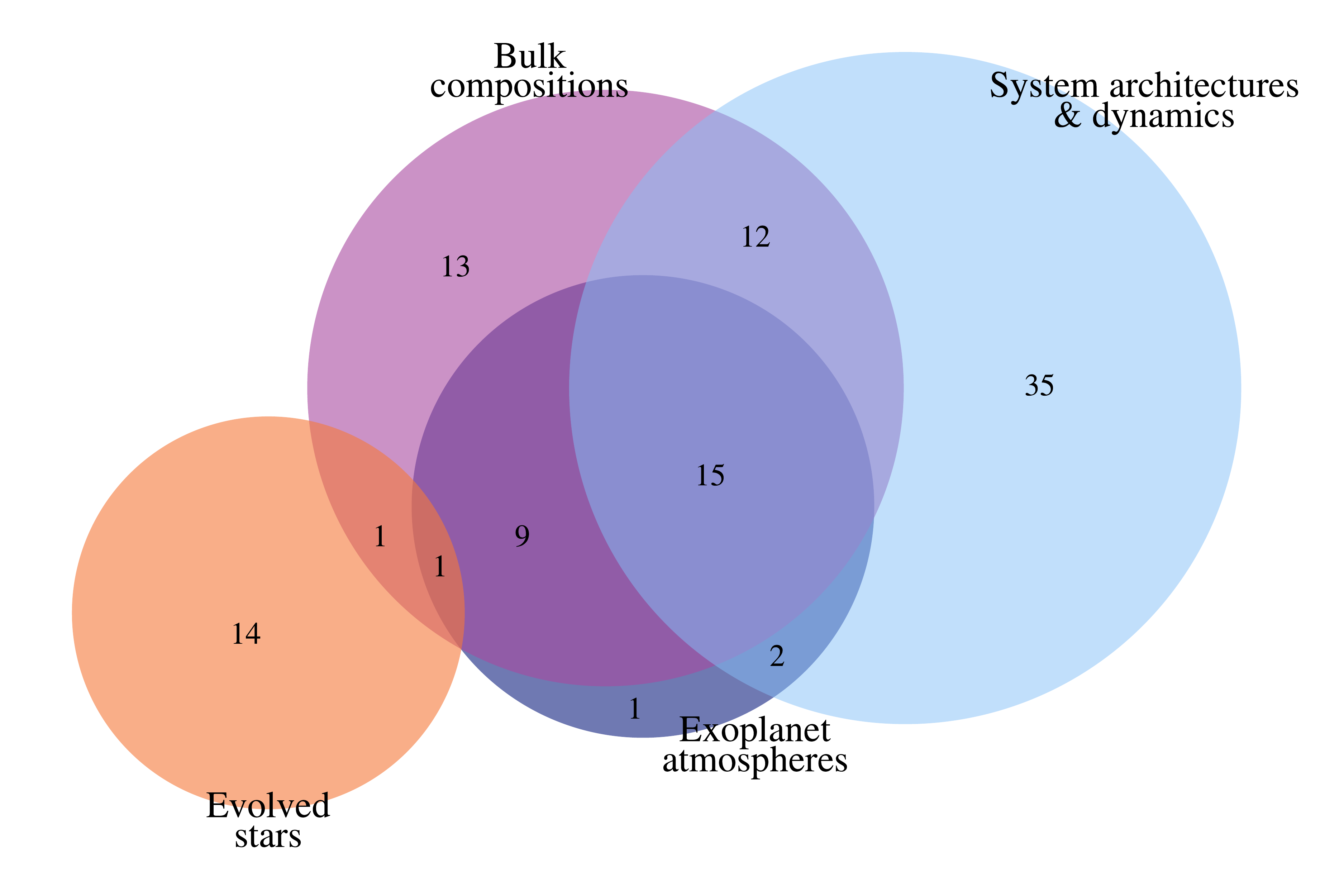}
\caption{Venn diagram of the selected planet sample divided into the four TKS science cases: 1) bulk compositions, 2) system architectures \& dynamics, 3) exoplanet atmospheres and 4) evolved stars. Numbers indicate the number of \textit{planets} belonging to each unique set of science cases, where total numbers include 51 planets for SC1, 64 planets for SC2, 28 planets for SC3, and 16 planets for SC4. The entire sample will be used for investigating any star-planet correlations (SC1E), characterizing properties of TKS host stars (TA), as well as exploring Doppler noise contributions and/or dependencies (TB).}
\label{fig:venn}
\end{figure*}

A data-flow diagram of the target prioritization algorithm is shown in Figure \ref{fig:flowselect}. In summary, the following steps occur while the total remaining time, $\mathrm{T_{tot}} \equiv \sum_{n} \mathrm{T}_{n}$ for $n$ science cases in a \texttt{Survey} is greater than zero:
\begin{enumerate}
    \item A program is selected at random, with a probability of selection that is proportional to the fractional amount of time remaining for each science case in the \texttt{Survey}. Specifically, the program calculates the cumulative distribution function by normalizing the most current list of remaining times for $n$ science cases by $\mathrm{T_{tot}}$. Therefore, a uniform random number on the unit interval $\sim[0,1]$ will accurately map back to the $n$ discrete science cases and select a program from the \texttt{Survey}.
    \item The vetted \texttt{Survey} sample (see Section \ref{sec:vet}) is filtered based on the selected program's selection criteria (see Section \ref{sec:scc}) to include only the relevant science targets. This is done by initializing a \texttt{Sample} class object, which updates the target \texttt{Sample} costs based on past algorithm selections. After cost updates are calculated, targets are ranked based on the prioritization metric(s) of the selected program. If the program provided any high priority targets, all priority targets would be ranked first, followed by the ordinal rankings.
    \item The selected program chooses the highest ranked star that has not yet been selected by that program. 
    \begin{enumerate}
        \item If the selected program \emph{can} afford the target (i.e. has the available time), the program is charged the required observing time for the selected star and the target is appended to the final \texttt{Survey} target list. If the target has already been selected by other programs in previous iterations, relevant programs are credited back any differences in costs.
        \item If the selected program \emph{cannot} afford the target (i.e. the required time for the highest ranked target exceeds the remaining time available for the program), the program is temporarily \texttt{stuck}. The ``stuck'' feature enabled the continuation of the algorithm by allowing a program to pass without removing it from the selection process altogether. The main reason for this is that as the algorithm progresses, including more target selections, a given target might suddenly become affordable if it is shared by enough programs.
    \end{enumerate}
\end{enumerate}

\noindent The above steps repeat until either 1) all resources have been exhausted (i.e.\texttt{$\rm \,\,T_{tot} = 0$}) or 2) all programs in the \texttt{Survey} are ``stuck''. Once one of these conditions is satisfied, the selection process is complete and the program will save a number of data products. The data products include a csv file with the target list and the \texttt{Track}, which records the detailed history (i.e. every iteration) of the selection process. Finally, an additional Monte Carlo-like simulation option is available to test how robust (or sensitive) the target sample is due to the inherent randomness of the selection process. 

\begin{deluxetable*}{rlrrcccrcl}
\tablecaption{TKS Target Sample}

\tablewidth{0pt}
\startdata
\vspace{-0.1cm}
& & & & & & & & & \\
\vspace{-0.05cm}
& & RA$\,\,$ & $\delta\,\,\,\,\,$ & V & \teff\ & \radiusp & Period & \sinc\ & \\
TOI & TIC & (deg) & (deg) & (mag) & (K) & (\radiuse) & (days) & (\since) & Science Cases$^{*}$ \\
\noalign{\smallskip}
\hline
\noalign{\smallskip}
260.01 & 37749396 & 4.7732 & $-$9.9648 & 9.90 & 4049 & 1.64 & 13.47 & 1.1e+01 & 1A, 1B, TA, TB$^{\dagger}$ \\ 
266.01 & 164767175 & 26.2096 & $-$18.4009 & 10.07 & 5784 & 2.41 & 10.77 & 1.0e+02 & 1A, 1B, 3$^{\dagger}$, TA, TB \\ 
266.02 & -- & -- & -- & -- & -- & 1.76 & 6.19 & 2.2e+02 & 1A, 1B, 3, TA, TB \\ 
329.01 & 169765334 & 351.3209 & $-$15.6347 & 11.26 & 5560 & 10.84 & 5.70 & 3.6e+03 & 4, TA, TB \\ 
465.01 & 270380593 & 32.7818 & $+$2.4180 & 11.56 & 4936 & 5.64 & 3.84 & 1.5e+02 & 2A, TA, TB \\ 
469.01 & 33692729 & 93.0582 & $-$14.6500 & 9.49 & 5283 & 3.69 & 13.63 & 8.9e+01 & 1B, 3$^{\dagger}$, TA, TB \\ 
480.01 & 317548889 & 88.3787 & $-$16.2650 & 7.28 & 6212 & 3.08 & 6.87 & 3.7e+02 & 1A, 1B, 4$^{\dagger}$, TA, TB \\ 
509.01 & 453211454 & 117.9250 & $+$9.3861 & 8.58 & 5560 & 3.07 & 18.12 & 4.6e+01 & 1A, 1B, 2A, 3$^{\dagger}$, TA, TB \\ 
554.01 & 407966340 & 60.7479 & $+$9.2085 & 6.91 & 6337 & 3.41 & 7.05 & 4.2e+02 & 1A, 1B, 3$^{\dagger}$, TA, TB \\ 
561.01 & 377064495 & 148.1856 & $+$6.2164 & 10.25 & 5440 & 3.77 & 10.78 & 6.6e+01 & 1A, 1B, 1C, 2C, 3, TA, TB \\ 
561.02 & -- & -- & -- & -- & -- & 1.55 & 0.45 & 4.6e+03 & 1A, 1B, 1C, 2C, 3$^{\dagger}$, TA, TB \\ 
561.03 & -- & -- & -- & -- & -- & 2.84 & 16.37 & 3.8e+01 & 1A, 1B, 1C, 2C, 3, TA, TB \\ 
669.01 & 124573851 & 158.9006 & $-$5.1817 & 10.61 & 5624 & 3.84 & 3.95 & 6.4e+02 & 1B, 3$^{\dagger}$, TA, TB \\ 
1136.01 & 142276270 & 192.1849 & $+$64.8553 & 9.53 & 5767 & 4.72 & 12.52 & 7.7e+01 & 1A, 1B, 2C, 3, TA, TB$^{\dagger}$ \\ 
1136.02 & -- & -- & -- & -- & -- & 3.00 & 6.26 & 2.1e+02 & 1A, 1B, 2C, 3, TA, TB$^{\dagger}$ \\ 
1136.03 & -- & -- & -- & -- & -- & 4.04 & 26.32 & 3.0e+01 & 1A, 1B, 2C, 3$^{\dagger}$, TA, TB$^{\dagger}$ \\ 
1136.04 & -- & -- & -- & -- & -- & 2.54 & 18.80 & 4.8e+01 & 1A, 1B, 2C, 3, TA, TB$^{\dagger}$ \\ 
1173.01 & 232967440 & 197.6823 & $+$70.7684 & 11.04 & 5322 & 9.22 & 7.06 & 8.1e+01 & 2A, TA, TB \\ 
1174.01 & 154089169 & 209.2181 & $+$68.6180 & 10.96 & 5077 & 2.31 & 8.95 & 4.4e+01 & 2A, TA, TB \\ 
1180.01 & 158002130 & 214.5531 & $+$82.1940 & 11.02 & 4700 & 2.85 & 9.69 & 3.5e+01 & 2A, TA, TB \\ 
1181.01 & 229510866 & 297.2159 & $+$64.3544 & 10.58 & 6122 & 16.66 & 2.10 & 4.0e+03 & 4, TA, TB \\ 
1184.01 & 233087860 & 272.2039 & $+$60.6782 & 10.99 & 4534 & 2.40 & 5.75 & 5.3e+01 & 2Bi$^{\dagger}$, TA, TB \\ 
1194.01 & 147950620 & 167.8205 & $+$69.9647 & 11.30 & 5323 & 8.86 & 2.31 & 6.4e+02 & 2A, TA, TB \\ 
1244.01 & 219850915 & 256.2803 & $+$69.5194 & 11.93 & 4599 & 2.39 & 6.40 & 5.6e+01 & 2A, TA, TB \\ 
1246.01 & 230127302 & 251.1165 & $+$70.4296 & 11.63 & 5141 & 3.33 & 18.65 & 2.7e+01 & 2A, 2C, TA, TB \\ 
1246.02 & -- & -- & -- & -- & -- & 3.00 & 4.31 & 2.0e+02 & 2A, 2C, TA, TB \\ 
1246.03 & -- & -- & -- & -- & -- & 2.63 & 5.90 & 1.3e+02 & 2A, 2C, TA, TB \\ 
1246.04 & -- & -- & -- & -- & -- & 3.26 & 37.92 & 1.0e+01 & 2A, 2C, TA, TB \\ 
1247.01 & 232540264 & 227.8705 & $+$71.8410 & 9.08 & 5711 & 2.80 & 15.92 & 7.0e+01 & 1A, 1B, 2A, 3$^{\dagger}$, TA, TB$^{\dagger}$ \\ 
1248.01 & 232612416 & 259.0236 & $+$63.1060 & 11.81 & 5227 & 6.62 & 4.36 & 2.0e+02 & 2A, TA, TB \\ 
1249.01 & 232976128 & 200.5617 & $+$66.3087 & 11.09 & 5453 & 3.15 & 13.08 & 3.1e+01 & 2A, TA, TB \\ 
1255.01 & 237222864 & 296.2443 & $+$74.0627 & 9.92 & 5126 & 2.70 & 10.29 & 5.9e+01 & 1A, 1B, 2A, 2Bi$^{\dagger}$, 3, TA, TB \\ 
1269.01 & 198241702 & 249.6971 & $+$64.5587 & 11.62 & 5517 & 2.38 & 4.25 & 2.3e+02 & 2A, TA, TB \\ 
1269.02 & -- & -- & -- & -- & -- & 2.32 & 9.24 & 8.2e+01 & 2A, TA, TB \\
\enddata
\tablecomments{\\
Values in this table represent what was used during the target selection and therefore, do not necessarily reflect the most up-to-date values. Duplicated target information for TOIs with more than one transiting planet candidate have been omitted. \\
$^{*}$ Please refer to Table \ref{tab:themes} for the definition of science case keys. \\
$^{\dagger}$ High-priority targets provided during the target selection process, as indicated by the individual program.}
\label{tab:sample1}
\end{deluxetable*}

\begin{deluxetable*}{rlrrcccrcl}
\tablecaption{TKS Target Sample (continued)}

\tablewidth{0pt}
\startdata
\vspace{-0.1cm}
& & & & & & & & & \\
\vspace{-0.05cm}
& & RA$\,\,$ & $\delta\,\,\,\,\,$ & V & \teff\ & \radiusp & Period & \sinc\ & \\
TOI & TIC & (deg) & (deg) & (mag) & (K) & (\radiuse) & (days) & (\since) & Science Cases$^{*}$ \\
\noalign{\smallskip}
\hline
\noalign{\smallskip}
1272.01 & 417948359 & 199.1966 & $+$49.8610 & 11.76 & 4987 & 4.29 & 3.32 & 2.2e+02 & 2A, 2Bi$^{\dagger}$, 3, TA, TB \\ 
1279.01 & 224297258 & 185.0639 & $+$56.2011 & 10.71 & 5477 & 2.58 & 9.62 & 1.0e+02 & 2A, TA, TB \\ 
1288.01 & 365733349 & 313.1666 & $+$65.6091 & 10.45 & 6180 & 4.73 & 2.70 & 6.9e+02 & 2A, TA, TB \\ 
1294.01 & 219015370 & 223.0929 & $+$70.4766 & 11.31 & 5714 & 9.30 & 3.92 & 3.2e+02 & 4, TA, TB \\ 
1296.01 & 219854185 & 256.7709 & $+$70.2385 & 11.37 & 5494 & 14.86 & 3.94 & 8.4e+02 & 4, TA, TB \\ 
1298.01 & 237104103 & 241.3234 & $+$70.1899 & 11.89 & 5731 & 10.15 & 4.54 & 5.1e+02 & 4, TA, TB \\ 
1339.01 & 269701147 & 302.0240 & $+$66.8506 & 8.97 & 5461 & 3.20 & 8.88 & 9.9e+01 & 1A, 1B, 2A, 2C, 3$^{\dagger}$, TA, TB \\ 
1339.02 & -- & -- & -- & -- & -- & 3.07 & 28.58 & 2.0e+01 & 1A, 1B, 2A, 2C, 3, TA, TB \\ 
1339.03 & -- & -- & -- & -- & -- & 1.74 & -- & 2.5e+01 & 1A, 1B, 2A, 2C, 3, TA, TB \\ 
1347.01 & 229747848 & 280.3268 & $+$70.2899 & 11.17 & 5424 & 2.06 & 0.85 & 1.9e+03 & 1A$^{\dagger}$, 1B, 1C, 2Bi$^{\dagger}$, TA, TB \\
1347.02 & -- & -- & -- & -- & -- & 1.78 & 4.84 & 1.8e+02 & 1A$^{\dagger}$, 1B, 1C, 2Bi, TA, TB \\ 
1386.01 & 343019899 & 334.5037 & $+$54.3190 & 10.61 & 5769 & 6.55 & -- & -1.0e+00 & 2Bi$^{\dagger}$, TA, TB \\ 
1410.01 & 199444169 & 334.8832 & $+$42.5603 & 11.11 & 4507 & 2.94 & 1.22 & 5.6e+02 & 1A, 1B, 2A, 3$^{\dagger}$, TA, TB \\ 
1411.01 & 116483514 & 232.9451 & $+$47.0568 & 10.51 & 4184 & 1.37 & 1.45 & 2.8e+02 & 2A, TA, TB \\ 
1422.01 & 333473672 & 354.2412 & $+$39.6394 & 10.62 & 5914 & 3.09 & 13.00 & 8.7e+01 & 2A, TA, TB \\ 
1430.01 & 293954617 & 300.6143 & $+$53.3768 & 9.19 & 5064 & 2.23 & 7.43 & 7.2e+01 & 1A, 1B, 3$^{\dagger}$, TA, TB \\ 
1436.01 & 154383539 & 215.6025 & $+$55.3337 & 12.01 & 5011 & 1.72 & 0.87 & 1.1e+03 & 1C$^{\dagger}$, 1E$^{\dagger}$, TA, TB$^{\dagger}$ \\ 
1437.01 & 198356533 & 256.1387 & $+$56.8426 & 9.17 & 6093 & 2.35 & 18.84 & 8.7e+01 & 1A, 1B, 2A, TA, TB \\ 
1438.01 & 229650439 & 280.9243 & $+$74.9376 & 10.96 & 5259 & 2.81 & 5.14 & 1.2e+01 & 2A, TA, TB \\ 
1438.02 & -- & -- & -- & -- & -- & 2.32 & 9.43 & 6.4e+01 & 2A, TA, TB \\ 
1439.01 & 232982558 & 241.7639 & $+$67.8777 & 10.55 & 5873 & 3.43 & 27.64 & 3.9e+01 & 4, TA, TB \\ 
1443.01 & 237232044 & 297.4003 & $+$76.1391 & 10.68 & 5236 & 2.07 & 23.54 & 1.5e+01 & 2A, TA, TB \\ 
1444.01 & 258514800 & 305.4743 & $+$70.9438 & 10.94 & 5466 & 1.29 & 0.47 & 5.0e+03 & 1C$^{\dagger}$, 2A, TA, TB \\ 
1451.01 & 417931607 & 186.5243 & $+$61.2590 & 9.58 & 5781 & 2.46 & 16.54 & 2.6e+01 & 1A, 1B, 2A, TA, TB \\ 
1456.01 & 199376584 & 306.7413 & $+$33.7445 & 8.56 & 6125 & 7.70 & 12.16 & 2.2e+02 & 3, TA, TB \\ 
1467.01 & 240968774 & 19.1139 & $+$49.2338 & 12.29 & 3834 & 1.83 & 5.97 & 1.8e+01 & 1E$^{\dagger}$, TA, TB \\ 
1471.01 & 306263608 & 30.9043 & $+$21.2809 & 9.20 & 5625 & 4.28 & 20.77 & 8.1e+01 & 2A, 3$^{\dagger}$, TA, TB \\ 
1472.01 & 306955329 & 14.1136 & $+$48.6376 & 11.30 & 5103 & 4.31 & 6.36 & 1.1e+02 & 2A, TA, TB \\ 
1473.01 & 352413427 & 15.5982 & $+$37.1855 & 8.84 & 5958 & 2.49 & 5.26 & 3.0e+02 & 1A, 1B, 3$^{\dagger}$, TA, TB$^{\dagger}$ \\ 
1601.01 & 139375960 & 38.3614 & $+$41.0134 & 10.66 & 5917 & 14.61 & 5.33 & 1.1e+03 & 4, TA, TB \\ 
1611.01 & 264678534 & 325.1866 & $+$84.3335 & 8.37 & 5071 & 2.72 & 16.19 & 6.6e+01 & 1A, 1B, 2A, TA, TB \\ 
1669.01 & 428679607 & 45.9529 & $+$83.5876 & 10.22 & 5550 & 2.25 & 2.68 & 6.9e+02 & 2A, TA, TB \\ 
1691.01 & 268334473 & 272.4061 & $+$86.8596 & 10.13 & 5759 & 3.75 & 16.73 & 5.9e+01 & 2A, TA, TB \\ 
1694.01 & 396740648 & 97.7482 & $+$66.3607 & 11.45 & 5058 & 5.48 & 3.77 & 2.0e+02 & 2A, TA, TB \\ 
\enddata
\tablecomments{\\
Values in this table represent what was used during the target selection and therefore, do not necessarily reflect the most up-to-date values. Duplicated target information for TOIs with more than one transiting planet candidate have been omitted. \\
$^{*}$ Please refer to Table \ref{tab:themes} for the definition of science case keys. \\
$^{\dagger}$ High-priority targets provided during the target selection process, as indicated by the individual program.}
\label{tab:sample2}
\end{deluxetable*}

\begin{deluxetable*}{rlrrcccrcl}
\tablecaption{TKS Target Sample (continued)}

\tablewidth{0pt}
\startdata
\vspace{-0.1cm}
& & & & & & & & & \\
\vspace{-0.05cm}
& & RA$\,\,$ & $\delta\,\,\,\,\,$ & V & \teff\ & \radiusp & Period & \sinc\ & \\
TOI & TIC & (deg) & (deg) & (mag) & (K) & (\radiuse) & (days) & (\since) & Science Cases$^{*}$ \\
\noalign{\smallskip}
\hline
\noalign{\smallskip}
1710.01 & 445805961 & 94.2828 & $+$76.2108 & 9.54 & 5675 & 5.41 & 24.28 & 3.5e+01 & 2A, TA, TB \\ 
1716.01 & 14336130 & 105.0829 & $+$56.8244 & 9.41 & 5878 & 2.74 & 8.09 & 2.5e+02 & 1A, 1B, 2A, TA, TB \\ 
1723.01 & 71431780 & 116.7971 & $+$68.4766 & 9.66 & 5777 & 3.16 & 13.72 & 9.1e+01 & 2A, TA, TB \\ 
1726.01 & 130181866 & 117.4794 & $+$27.3632 & 6.92 & 5694 & 2.16 & 7.11 & 1.4e+02 & 1A, 1B, TA, TB$^{\dagger}$ \\ 
1726.02 & -- & -- & -- & -- & -- & 2.64 & 20.55 & 3.5e+01 & 1A, 1B, 2Bii$^{\dagger}$, TA, TB$^{\dagger}$ \\ 
1736.01 & 408618999 & 43.4350 & $+$69.1014 & 8.95 & 5656 & 2.74 & 7.07 & 2.0e+02 & 1A, 1B, 3$^{\dagger}$, 4, TA, TB \\ 
1742.01 & 219857012 & 257.3285 & $+$71.8764 & 8.86 & 5707 & 2.20 & 21.27 & 5.5e+01 & 1A, 1B, 2A, TA, TB \\ 
1751.01 & 287080092 & 243.4888 & $+$63.5343 & 9.33 & 6114 & 2.81 & 37.47 & 3.8e+01 & 1A, 1B, 2A, TA, TB \\ 
1753.01 & 289580577 & 252.4698 & $+$61.1735 & 11.83 & 5700 & 2.96 & 5.38 & 2.4e+02 & 2A, TA, TB \\ 
1758.01 & 367858035 & 354.7430 & $+$75.6851 & 10.79 & 5169 & 3.80 & 20.70 & 2.0e+01 & 2A, TA, TB \\ 
1759.01 & 408636441 & 326.8533 & $+$62.7539 & 11.93 & 3960 & 3.23 & 37.70 & 2.0e+00 & 1D$^{\dagger}$, 2A, 3$^{\dagger}$, TA, TB \\ 
1775.01 & 9348006 & 150.1151 & $+$39.4578 & 11.65 & 5251 & 8.07 & 10.24 & 6.0e+01 & 2A, TA, TB \\ 
1776.01 & 21535395 & 164.7761 & $+$40.9836 & 8.26 & 5723 & 1.40 & 2.80 & 5.6e+02 & 1A, 1B, 3, TA, TB \\ 
1778.01 & 39699648 & 136.7781 & $+$46.6726 & 8.99 & 6023 & 2.83 & 6.52 & 4.1e+02 & 1A, 1B, TA, TB \\ 
1794.01 & 286916251 & 203.3977 & $+$49.0611 & 10.32 & 5707 & 3.03 & 8.77 & 2.0e+02 & 2A, TA, TB \\ 
1797.01 & 368435330 & 162.7771 & $+$25.6412 & 9.18 & 5922 & 3.21 & 3.65 & 5.4e+02 & 1A, 1B, 2A, TA, TB \\ 
1798.01 & 198153540 & 211.0941 & $+$46.5194 & 11.36 & 5165 & 2.39 & 8.02 & 6.9e+01 & 1C$^{\dagger}$, TA, TB \\ 
1798.02 & -- & -- & -- & -- & -- & 1.28 & 0.44 & 3.4e+03 & 1C, TA, TB \\ 
1799.01 & 8967242 & 167.2330 & $+$34.3032 & 8.98 & 5690 & 1.63 & 7.09 & 1.6e+02 & 1A, 1B, TA, TB \\ 
1801.01 & 119584412 & 175.5766 & $+$23.0269 & 11.58 & 3815 & 2.17 & 21.28 & 3.0e+00 & 1D$^{\dagger}$, 1E$^{\dagger}$, TA, TB \\ 
1807.01 & 180695581 & 201.2833 & $+$38.9225 & 10.00 & 4612 & 1.53 & 0.55 & 1.6e+03 & 1A, 1C, TA, TB$^{\dagger}$ \\ 
1823.01 & 142381532 & 196.2204 & $+$63.7538 & 10.73 & 4760 & 8.14 & 194.05 & 9.0e+00 & 2A, TA, TB \\ 
1824.01 & 142387023 & 197.7312 & $+$61.7448 & 9.72 & 5182 & 2.40 & 22.81 & 1.7e+01 & 1A, 1B, 2A, TA, TB$^{\dagger}$ \\ 
1836.01 & 207468071 & 245.9082 & $+$54.6898 & 9.77 & 6351 & 7.88 & 20.38 & 1.1e+02 & 4, TA, TB \\ 
1842.01 & 404505029 & 201.9628 & $+$9.0307 & 9.81 & 6115 & 12.68 & 19.15 & 2.4e+02 & 4, TA, TB \\ 
1898.01 & 91987762 & 144.5556 & $+$23.5469 & 7.87 & 6303 & 7.16 & -- & 2.0e+02 & 4, TA, TB \\ 
1905.01 & 429302040 & 188.3869 & $-$10.1461 & 11.59 & 4251 & 12.84 & 5.72 & 3.7e+01 & 2Bii, TA, TB \\ 
2019.01 & 159781361 & 234.4317 & $+$48.9554 & 10.26 & 5588 & 6.09 & 15.35 & 1.9e+02 & 4, TA, TB \\ 
2045.01 & 347013211 & 1.1191 & $+$54.9345 & 11.30 & 6125 & 12.97 & 9.08 & 5.4e+02 & 4, TA, TB \\ 
2076.01 & 27491137 & 217.3927 & $+$39.7904 & 9.14 & 5163 & 2.89 & 10.36 & 4.8e+01 & 1A, 1B, 1D$^{\dagger}$, TA, TB \\ 
2076.02 & -- & -- & -- & -- & -- & 4.30 & 33.69 & 9.0e+00 & 1A, 1B, 1D$^{\dagger}$, TA, TB \\ 
2088.01 & 441765914 & 261.3752 & $+$75.8823 & 11.64 & 4902 & 3.51 & 124.73 & 1.0e+00 & 1D$^{\dagger}$, 1E$^{\dagger}$, 3, TA, TB \\ 
2114.01 & 9828416 & 261.0964 & $+$33.2051 & 10.27 & 6382 & 13.87 & 6.21 & 5.3e+02 & 4, TA, TB \\ 
2128.01 & 21832928 & 256.9826 & $+$32.1055 & 7.22 & 5991 & 1.97 & 16.33 & 8.6e+01 & 1A, 1B, TA, TB \\ 
2145.01 & 88992642 & 263.7581 & $+$40.6951 & 9.07 & 6202 & 12.41 & 10.26 & 1.1e+03 & 4, TA, TB \\ 
\enddata
\tablecomments{\\
Values in this table represent what was used during the target selection and therefore, do not necessarily reflect the most up-to-date values. Duplicated target information for TOIs with more than one transiting planet candidate have been omitted. \\
$^{*}$ Please refer to Table \ref{tab:themes} for the definition of science case keys. \\
$^{\dagger}$ High-priority targets provided during the target selection process, as indicated by the individual program.}
\label{tab:sample3}
\end{deluxetable*}

\section{TKS Target Sample} \label{sec:sample}

For TKS allocations, most science cases started with an equal amount of time (10\% of $\rm T_{tot}$), with the exception of SC1D that started with half of that allocation (5\% of $\rm T_{tot}$). In addition, 10\% of $\rm T_{tot}$ was designated to TB for high-cadence high-priority activity targets, while TA did not require any allocation. Figure \ref{fig:sample} shows the complete TKS sample (\plotcolor) selected by the prioritization algorithm using an allocation of 50 nights and assuming 10 hours per night. Almost all of the brighter targets available for selection were almost always selected (Figure \ref{fig:sample}a), which was an inherent product set by a characteristic of the algorithm (i.e. brighter targets are cheaper and therefore, typically more highly ranked). Moreover, most TKS targets are brighter than $V=12$, with the exception of one or two hand-selected targets.

\begin{figure*}[ht!]
\centering
\includegraphics[width=\linewidth]{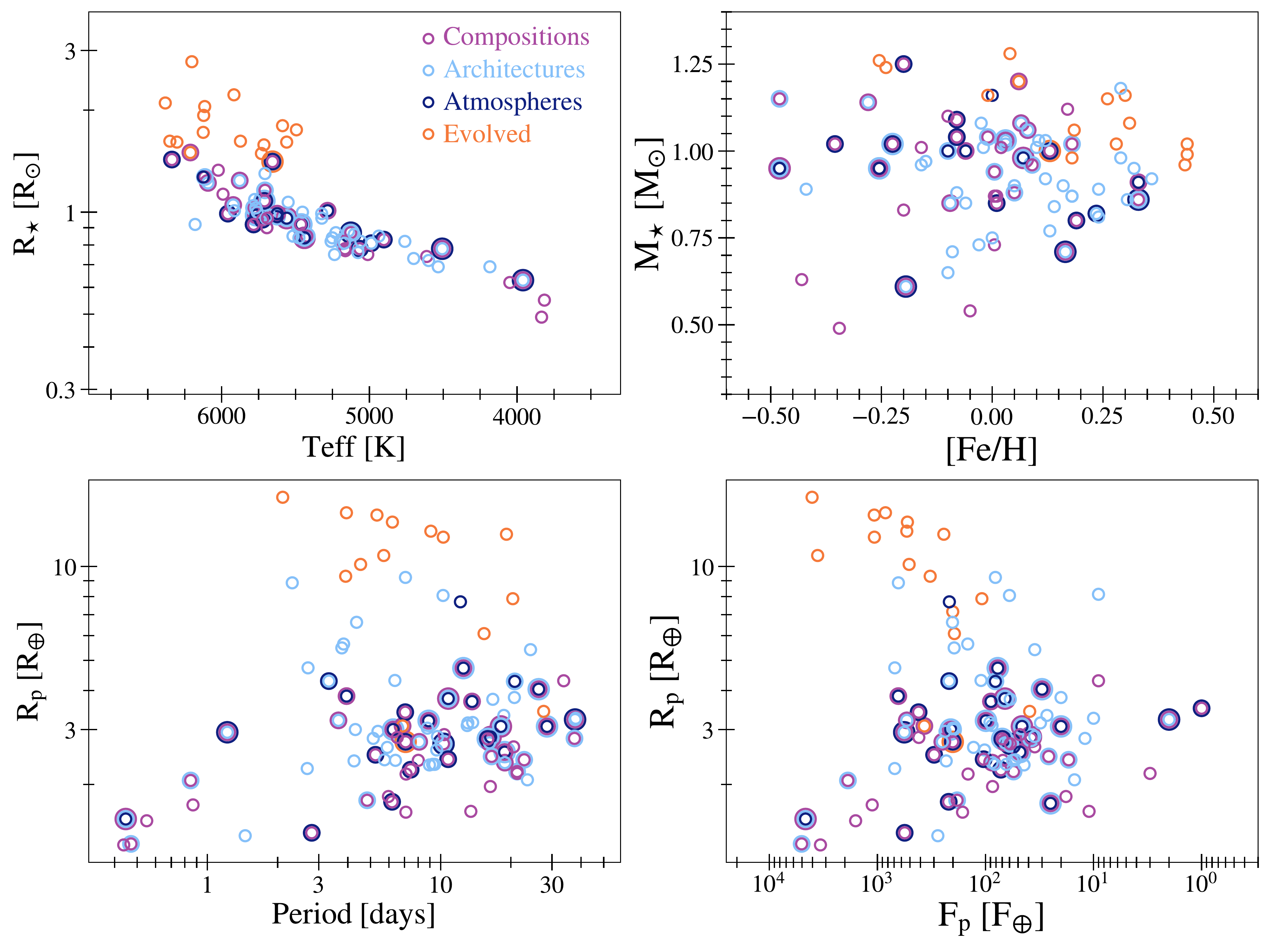}
\caption{Same as Figures \ref{fig:hosts} and \ref{fig:planets} but now only showing the selected TKS sample. Markers are colored by the four TKS science cases, where concentric rings represent higher-priority targets that address multiple science cases.}
\label{fig:programs}
\end{figure*}

Figure \ref{fig:hosts} shows various host star properties for the TKS sample. Of the selected \nhosts\ targets, 51 are solar-type stars, which is defined here as any star with an effective temperature within $\pm\,500$ K of the Sun. There are very few hot stars, most likely due to the rapid rotation of stars past the Kraft break (occurring near $\sim6250\,$ K). Stellar masses and metallicites appear clustered around solar-like values (Figure \ref{fig:hosts}, right), as expected from the effective temperature distribution. Fortuitously, the diverse range in metallicity was a natural outcome of the selection process. On the other hand, cool low-mass stars were selected less since they are fainter and therefore more expensive. Therefore in order to partially avoid this bias, TOI 1467 and TOI 1801 were added as high priority targets for SC1E and ended up being the two coolest stars in the TKS sample (with \teff\ $\leq 4000\,$K).

Figure \ref{fig:planets} shows the sample of \nplanets\ planets in \nhosts\ systems selected for the TESS-Keck Survey. Planet properties are slightly biased towards sub-Jovian planets, which was expected since all but TKS Evolved (SC4) preferentially selected smaller planets. In fact, the sample includes 71 planets smaller than $\leq4$ \radiuse\ that span nearly 5 orders of magnitude in incident flux, including a few that experience a similar amount to that received on Earth. Targets for the habitable zone science case are cooler than the majority of small planets with measured densities and therefore useful for exploring how planet properties change with decreasing incident flux. The TKS sample also includes a handful of planets at the opposite extreme, on sub-day orbital periods presumably with little to no remaining atmosphere.


While not unexpected, the strong overlap between the broad TKS science themes is remarkable. Figure \ref{fig:venn} shows a Venn diagram of the planet sample by science case, and Figure \ref{fig:programs} shows the samples in host star and planet parameter spaces. The full TKS sample demonstrates the proof of concept for the unique target selection process. Most notably, the large number of interdisciplinary targets that were automatically identified through the algorithm emphasizes the value for applying such target selection technique. This demonstrates the utility for other large surveys in the future that face similar challenges related to target selection processes, especially for collaborations with sub-teams and/or overlapping science cases.

The complete TKS sample selected by the algorithm is provided in Tables \ref{tab:sample1}-\ref{tab:sample3}, including the science case(s) for each target. Most notably, the highest priority targets within the TKS sample (i.e. targets prioritized by the most science cases) have been or will be highlighted by early TKS single-object papers. An example is the galactic thick disc multi-planet system with an ultra-short-period planet, TOI 561 \citep{weiss2021}. Other single-system TKS papers nearing publication include the high-eccentricity candidate TOI 1255 (MacDougall et al., in prep.), four transiting sub-Neptunes with diverse masses around TOI 1246 (Turtelboom et al., in prep.), a pair of prime atmospheric, sub-Neptune twins orbiting HD 63935 (Scarsdale et al., in prep.), as well as the five-planet system HD 191939 (Lubin et al., in prep). The entire TKS sample provided in this paper has not been modified since August 21st, 2020, but is still subject to change as the survey continues.

\section{Conclusions}

In this paper, we have presented the science cases and target selection process for the TESS-Keck Survey, a large program using Keck/HIRES to confirm and characterize planets discovered with TESS. 

Our main conclusions are as follows:
\begin{itemize}
\item The TESS-Keck survey will measure precise ($>5\sigma$) masses for $\sim100$ planets using an allocation of $\sim$100 nights over four semesters. TKS will leverage this new population of transiting exoplanets orbiting bright, nearby stars to address four main science themes, including 1) the bulk compositions of small planets, 2) dynamical temperatures and system architectures, 3) a larger, more refined sample for future atmospheric studies and 4) planets orbiting evolved stars (Section \ref{sec:science}).
\item We have developed open-source software for a fully-automated and reproducible target selection procedure, which can be adapted for and used by other surveys (Section \ref{sec:algo}). By providing a set of science programs with unique selection criteria and prioritization metrics, the ranking algorithm will randomly select programs and program targets until the allocated resources are exhausted.
\item A total of \nhosts\ targets were selected by the prioritization algorithm for the final TKS sample (Section \ref{sec:sample}). The majority of TKS hosts are brighter ($V<12$), solar-like main-sequence stars with effective temperatures, $\rm 4500 \,\,K \leq$ \teff\ $\rm <6000\,\,K$, at a wide range of metallicities ($-0.5 <$ \feh\ $< 0.5$). With the exceptions of hand-selected, high-priority targets for the obliquity and stellar activity programs, most stars have modest rotation and activity levels (Figures \ref{fig:sample}, \ref{fig:hosts}, and \ref{fig:programs}). 
\item The final TKS planet population comprises \nplanets\ transiting planets in a rich diversity of system configurations (Section \ref{sec:sample}, Tables \ref{tab:sample1}-\ref{tab:sample3}). The selected sample has $71$ small planets (\radiusp\ $\leq4$ \radiuse) with incident fluxes that span nearly 5 orders of magnitude, including a few that are in or near the habitable zone of their host star (Figures \ref{fig:sample}, \ref{fig:planets} and \ref{fig:programs}).
\end{itemize}

The target selection presented here is the first in a series of papers presenting ensemble results from the TESS-Keck Survey. Future planned catalogs include homogeneous stellar properties, exoplanet masses and dynamical architectures of planetary systems observed by the TESS Mission. The target prioritization algorithm presented in this paper is publicly available\footnote{\url{https://github.com/ashleychontos/sort-a-survey/}}, which is currently designed to reproduce the enclosed TKS target sample. However, the selection algorithm can be generalized and applicable to other large collaborations or surveys that require a balance of multiple science interests within a given allocation.

\section*{Acknowledgements}

We thank all the observers who spent time collecting data over the many years at Keck/HIRES. We are grateful to the time assignment committees of the University of California, University of Hawai'i, the California Institute of Technology, and NASA for supporting the TESS-Keck Survey with observing time at Keck Observatory and on the Automated Planet Finder.  We thank NASA for funding associated with our Key Strategic Mission Support project.  We gratefully acknowledge the efforts and dedication of the Keck Observatory staff for support of HIRES and remote observing.  We recognize and acknowledge the cultural role and reverence that the summit of Maunakea has within the indigenous Hawaiian community. We are deeply grateful to have the opportunity to conduct observations from this mountain.  We thank Ken and Gloria Levy, who supported the construction of the Levy Spectrometer on the Automated Planet Finder. We thank the University of California and Google for supporting Lick Observatory and the UCO staff for their dedicated work scheduling and operating the telescopes of Lick Observatory.  This paper is based on data collected by the TESS mission. Funding for the TESS mission is provided by the NASA Explorer Program.

A.C., J.M.A.M., R.A.R., A.B., and A.M. acknowledge support from the National Science Foundation through the Graduate Research Fellowship Program (DGE 1842402, DGE 1842400, DGE 1745301, DGE 1745301, DGE 1752814). J.M.A.M. also acknowledges the LSSTC Data Science Fellowship Program, which is funded by LSSTC, NSF Cybertraining Grant No. 1829740, the Brinson Foundation, and the Moore Foundation; his participation in the program has benefited this work. D.H. acknowledges support from the Alfred P. Sloan Foundation, the National Aeronautics and Space Administration (80NSSC18K1585, 80NSSC19K0379), and the National Science Foundation (AST-1717000). I.J.M.C. acknowledges support from the NSF through grant AST-1824644. C.D.D. acknowledges the support of the Hellman Family Faculty Fund, the Alfred P. Sloan Foundation, the David \& Lucile Packard Foundation, and the National Aeronautics and Space Administration via the TESS Guest Investigator Program (80NSSC18K1583). E.A.P. acknowledges the support of the Alfred P. Sloan Foundation. L.M.W. is supported by the Beatrice Watson Parrent Fellowship and NASA ADAP Grant 80NSSC19K0597. P.D. acknowledges support from a National Science Foundation Astronomy and Astrophysics Postdoctoral Fellowship under award AST-1903811. S.G. acknowledges support from NASA FINESST Grant 80NSSC20K1549.

\facilities{\href{https://archive.stsci.edu/index.html}{MAST}, TESS, Keck I: HIRES}

\software{All code used in this paper is available at \url{https://github.com/ashleychontos/sort-a-survey}. We made use of the following publicly-available Python modules: \texttt{astroquery}, \texttt{exoplanet} \citep{fm2019}, \texttt{evolstate} \citep{huber2017,berger2018}, and \texttt{tesspoint}.}

\vspace{1cm}
\small
\bibliography{main.bib}

\end{document}